\documentclass{article}

\usepackage{arxiv}

\usepackage[utf8]{inputenc} 
\usepackage[T1]{fontenc}    
\usepackage{hyperref}       
\usepackage{url}            
\usepackage{booktabs}       
\usepackage{amsfonts}       
\usepackage{nicefrac}       
\usepackage{microtype}      
\usepackage{graphicx}
\usepackage[numbers]{natbib}
\usepackage{doi}
\usepackage{subcaption}
\usepackage{floatrow}
\usepackage{siunitx}
\usepackage{amsmath}

\title{Short pyridine-furan springs exhibit bistable dynamics of Duffing oscillators}

\author{\href{https://orcid.org/0000-0002-2516-8868}{\includegraphics[scale=0.06]{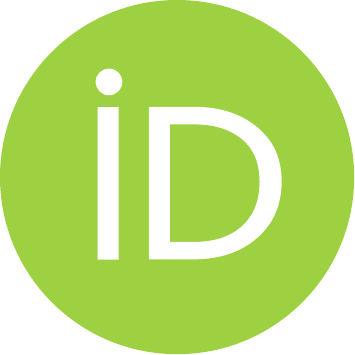}\hspace{1mm}Vladik A.~Avetisov} \\
	N. N. Semenov Federal Research Center of Chemical Physics of the Russian Academy of Sciences\\
	Kosygina 4\\
	119991 Moscow, Russia \\
    \AND
	Maria A.~Frolkina \\
	N. N. Semenov Federal Research Center of Chemical Physics of the Russian Academy of Sciences\\
	Kosygina 4\\
	119991 Moscow, Russia \\
	\AND
	\href{https://orcid.org/0000-0003-0699-7238}{\includegraphics[scale=0.06]{orcid.pdf}\hspace{1mm}Anastasia A.~Markina} \\
	N. N. Semenov Federal Research Center of Chemical Physics of the Russian Academy of Sciences\\
	Kosygina 4\\
	119991 Moscow, Russia \\
	\And
	
	\href{https://orcid.org/0000-0002-6769-0486}{\includegraphics[scale=0.06]{orcid.pdf}\hspace{1mm}Alexander D.~Muratov} \\
	N. N. Semenov Federal Research Center of Chemical Physics of the Russian Academy of Sciences\\
	Kosygina 4\\
	119991 Moscow, Russia \\
	\And
	\href{https://orcid.org/0000-0001-8913-8521}{\includegraphics[scale=0.06]{orcid.pdf}\hspace{1mm}Vladislav S.~Petrovskii}
	\\
	N. N. Semenov Federal Research Center of Chemical Physics of the Russian Academy of Sciences\\
	Kosygina 4\\
	119991 Moscow, Russia \\
}

\renewcommand{\shorttitle}{\textit{arXiv} Template}

\hypersetup{
pdftitle={Short pyridine-furan springs exhibit bistable dynamics of Duffing oscillators},
pdfsubject={q-bio.NC, q-bio.QM},
pdfauthor={Vladik A.~Avetisov, Maria A.~Frolkina, Anastasia A.~Markina, Alexander D.~Muratov, Vladislav Petrovskii},
pdfkeywords={First keyword, Second keyword, More},
}

\begin{document}
\maketitle

\begin{abstract}
The intensive development of nanodevices acting as two-state systems has motivated the search for nanoscale molecular structures whose dynamics are similar to those of bistable mechanical systems, such as Euler arches and Duffing oscillators.  Of particular interest are the molecular structures capable of spontaneous vibrations and stochastic resonance. Recently, oligomeric molecules that were a few nanometers in size and exhibited the bistable dynamics of an Euler arch were identified through molecular dynamics simulations of short fragments of thermo-responsive polymers subject to force loading. In this article, we present molecular dynamics simulations of short pyridine-furan springs a few nanometers in size and demonstrated the bistable dynamics of a Duffing oscillator with thermally-activated spontaneous vibrations and stochastic resonance.
\end{abstract}

\section{Introduction}
\label{sec:intro}
Nanoscale molecular structures, whose long-term dynamics resemble those of bistable mechanical systems, have been attracting more and more attention due to the intensive design and practical implementation of a wide range of nanodevices acting as switches and logic gates\cite{mi6081046,C5SC02317C,C7CS00491E,benda_substrate-dependent_2019,berselli_robust_2021,NICOLI2021213589}, sensors and actuators\cite{zhang_molecular_2018,shu_stimuli-responsive_2020,lemme_nanoelectromechanical_2020,shi_driving_2020,aprahamian_future_2020}, mechanoelectric transductors and energy harvesters\cite{li_energy_2014,kim_harvesting_2015,ackerman_anomalous_2016,dutreix_two-level_2020,thibado_fluctuation-induced_2020}. Nanoscale bistable systems are also no less important for verifying the foundation of stochastic thermodynamics \cite{evans_fluctuation_2002,seifert_stochastic_2012,horowitz_thermodynamic_2020,ciliberto_experiments_2017}, which is presently undergoing an extension of the thermodynamic theory as applied to nanoscale molecular machines\cite{ciliberto_experiments_2017,wang_experimental_2002,jop_work_2008,astumian_stochastic_2018,vroylandt_efficiency_2020}.

Two types of bistable mechanical systems can be considered as prototypes of nanoscale molecular structures for which this article is addressed. These are an Euler arch\cite{arnold_catastrophe_1984,poston_catastrophe_1996}, which looks like an elastic rod, and a Duffing oscillator\cite{duffing1918erzwungene,korsch_duffing_2008}, which is a spring with nonlinear elasticity. Both prototypes can be considered one-dimensional (1D) dynamic systems with critical behavior exhibiting bistability over a particular range of force loading. For example, an elastic rod slightly compressed in the longitudinal direction will remain straight. However, as soon as the compressive force exceeds a critical value, the straightened state becomes unstable and bifurcates into two arcuate states associated with the rod.  In energy terms, a potential energy function with a single minimum corresponding to the stability of the straightened state governs the dynamics of a sub-critically compressed Euler arch. In contrast, the potential of an Euler arch subject to super-critical compression has two energy wells corresponding to two symmetric arcuate states, which are separated from each other by the bistability barrier. Accordingly, switch-like transitions between the two states can be driven with lateral pushing of the arch. 

Besides deterministic transitions between the two states controlled by force loads, spontaneous vibrations, i.e., spontaneous jumps between these states, can be activated by noisy-like random disturbance of the bistable system. In the spontaneous vibration mode, the time intervals separating spontaneous jumps (the lifetimes of the system in its states) are random values distributed around an average lifetime, which exponentially grows with the ratio of the bistability barrier to the noise intensity in accordance with Kramer's rate approximation \cite{KRAMERS1940284}. Spontaneous vibrations are observed when this ratio is not too large (for instance, the bistability barrier is an order of magnitude greater than the noise intensity). In turn, spontaneous vibrations can be transformed into almost regular, but still noise-induced switching between the two states by slight wiggling of the bistable potential via weak oscillating force. This phenomenon was called stochastic resonance\cite{benzi_mechanism_1981}. Along with spontaneous vibrations, stochastic resonance is the most striking manifestation of bistability.  

In fact, stochastic resonance is a very peculiar combination of the non-linear dynamics of the system and its stochastic perturbations, with which the noise amplifies a weak signal rather than blurring it. Although the pioneering idea on stochastic resonance had been addressed to theoretical reasoning about the regularity of the ice ages on the Earth\cite{benzi_mechanism_1981,benzi_stochastic_1982,benzi_theory_1983}, it could not but cause an avalanche of publications devoted to the practical use and interpretations of stochastic resonance in a wide range of macroscopic, global, and even space systems\cite{gammaitoni_stochastic_1998,wellens_stochastic_2004}. To date, some experimental evidence has been obtained for which the bistable patterns might be present in scales down to sub-micron, for instance, in nanotubes\cite{baughman1999carbon, fujii2017single,huang_nonlocal_2019}, graphene sheets\cite{ackerman_anomalous_2016,liang2012electromechanical} , DNA hairpins, and proteins\cite{forns2011improving,hayashi2012single,cecconi2005direct}. In this regard, it should be noted that spontaneous vibrations and stochastic resonance of macroscopic mechanical systems, even if they are a micron in size, could hardly be activated by environmental thermal noise. The bistability barriers of macroscopic systems are much higher than the thermal noise intensity  ($\sim k_{B}T$); much stronger perturbations are required to activate the spontaneous vibrations and stochastic resonance of the mechanics even on a micron-scale.

However, nanoscale mechanics may provide a solution since the bistability barrier of a bistable nanoscale system may be high enough to separate two states of the system against thermal noise, and the same barrier may be low enough to allow the activation of the transitions by thermal-bath fluctuations. A value of about ten for the ratio of the bistability barrier to the noise intensity might serve as a reasonable reference point. Some oligomeric molecules within a few nanometers could be assumed to represent such bistable systems. Indeed, bistable molecules demonstrating the dynamics of an Euler arch was recently found through intensive molecular dynamic simulations of short thermo-responsive oligomeric compounds that were subject to force loads\cite{avetisov2019oligomeric,markina2020detection}. The simulations showed mechanic-like bistability of specific oligomeric molecules with spontaneous vibrations and stochastic resonance that were activated by thermal fluctuations.    

In this article, we present molecular dynamic simulations of short pyridine-furan springs as a continuation of the search for nanoscale molecular structures that exhibit bistability. Pyridine–furan (PF) springs attracted our attention since they could demonstrate non-linear dynamics due to nonlinear elasticity caused by the $\pi-\pi$ interactions between aromatic groups located on the adjacent turns of the spring. It is questionable as to whether an Euler arch could be considered a mechanical prototype of a nanoscale spring, whereas the Duffing oscillator can be viewed as the prototype.

In general, the Duffing oscillators form a class of non-linear dynamic systems specified by damped oscillations of springs with non-linear elasticity (see Section \ref{sec:MatMet1} for more details)\cite{duffing1918erzwungene}. In mechanics, ingenious combinations of springs were designed to implement a bistable Duffing oscillator (for an example see \citet{LAI201660,LU2021249}), so the search for a Duffing oscillator among nanoscale molecules may seem to be an unrealistically daunting task. However, the computer simulations presented below show that such a task is not really hopeless. Our studies of short PF springs support the idea that oligomeric springs with soft low-energy coupling of the turns due to $\pi-\pi$ stacking can possess the bistable dynamic characteristics of a Duffing oscillator. Moreover, beside the two-state deterministic dynamic, an oligomeric Duffing oscillator can exhibit thermally-activated spontaneous vibrations and stochastic resonance.

\section{Matherials and Methods}
\label{sec:MatMet}

\subsection{Duffing oscillators}\label{sec:MatMet1}
Deterministic dynamics of a Duffing oscillator obey the following Newtonian equation: 
\begin{equation}
\label{eq:1}
\frac{{d}^2 x}{{dt}^2}+k\frac{dx}{dt}=-\frac{dU(x)}{dx}
\end{equation}
in which $x$ is the deviation of a unit mass from the position $x=0$ (thereafter, this position is called a median zero-stress point), $k$ is a damping parameter, and $U(x)=ax^2+bx^4, b>0$is a four-degree potential of a spring, assuming that the spring elasticity changes linearly at small deviations, $x$, while it increases non-linearly at large deviations, $x$. It is not difficult to see that given the positive elasticity coefficient, $a$, the potential $U(x)$ has single extremum (minimum) located at the median zero-stress point $x=0$, so the spring experiences damped oscillations around this point. At large deviations from the median zero-stress point, the nonlinear effects, such as non-isochronism and anharmonicity, may accompany the oscillations\cite{korsch_duffing_2008}. However, the potential $U(x)$ becomes bistable if the linear elasticity coefficient, $a$, is negative. In this case, the spring has three zero-stress points ($x_{1}=0, x_{2,3}=\pm ({-\nicefrac{a}{2b}})^{\nicefrac{1}{2}}$). Two of them ($x_2$ and $x_3$) specify the attraction basins located at large deviations from the median zero-stress point while this point becomes unstable and repulses the dynamic trajectories. Therefore, depending on the values of the parameters $k$, $a$, and $b$, the Duffing oscillator either has one stable attractor in the form of a node or a spiral point, or it is bistable and has two attractive nodes or spiral points. To be bistable, the spring should have elasticity that causes a decrease in the elastic energy with small deviations from the median zero-stress point and increases the elastic energy with large deviations, $x$.

Bistable springs manifest the best behavior when random perturbations and oscillating forces are applied to them. In such cases, the dynamics of bistable springs are described by the Langevin equation of the form:

\begin{equation}
    \label{eq:2}
    \frac{{d}^2 x}{{dt}^2}+k\frac{dx}{dt}=-2ax-4bx^3+2 \epsilon f(t) +E_0 \cos(\omega t),
\end{equation}
where $a$ is negative and $b$ is positive, $f(t)$ denotes a zero-mean, Gaussian white noise with autocorrelation function $\langle f(t),f(0)\rangle=\delta(t)$, and the last term in the right-hand part of equation \ref{eq:2} is an external field oscillating with a circular frequency, $\omega$.

The interest in the action of random perturbations and oscillating field on bistable springs arose because random perturbations could activate random transitions between the two attracting basins of the spring while the oscillating field can force the regular transitions between the attracting basins. As a result, the dynamics of a nonlinear spring turn out to be multimode in contrast to the dynamics of a linear spring. In addition to deterministic behavior associated with non- or weekly-dumped oscillations in a single attracting basin, bistable spring can exhibit spontaneous vibrations and stochastic resonance caused by random and forced transitions between the two basins, respectively\cite{korsch_duffing_2008}. The implementation of bistability, spontaneous vibrations, and stochastic resonance using nano-sized springs immersed in the thermal bath as the only source of random perturbations could be of the greatest interest.

\subsection{Pyridine-furan springs}\label{sec:MatMet2}

\begin{figure}
    \centering
    \includegraphics[width=\linewidth]{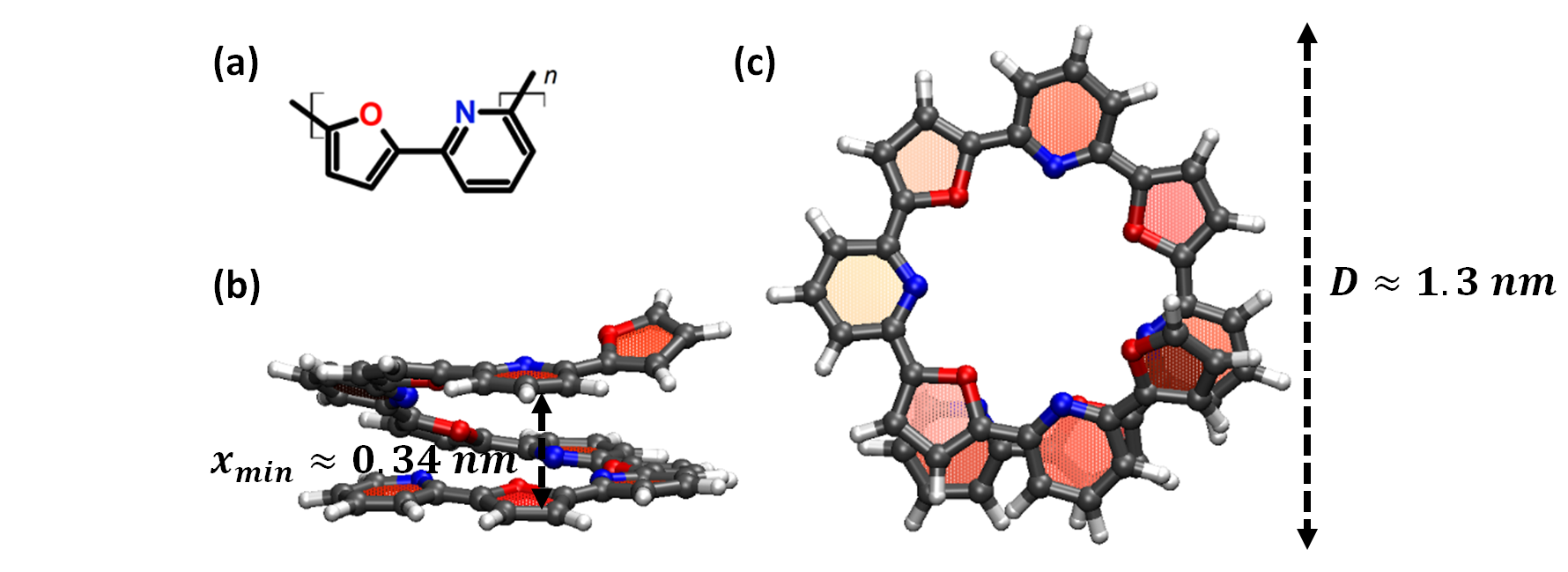}
    \caption{Pyridine–furan (PF) spring with five monomer units (oligo-PF-$5$ spring): (a) Chemical structure of a pyridine-furan monomer unit with heterocyclic rings in cis-configuration. (b) Front and (c) top views of an oligo-PF-$5$ spring in the atomistic representation. The spring has one complete turn consisting of approximately $3.5$ monomer units.}
    \label{fig:fig1}
\end{figure}
A PF copolymer (Figure \ref{fig:fig1}a)) is a conductive polymer consisting of $5$- and $6$-member heterocyclic rings as synthesized and described by \citet{jones1997extended}. PF copolymers tend to assume a helix-like shape, which is squeezed by the $\pi -\pi$ interactions of aromatic groups located at the adjacent turns\cite{sahu2015}. Assuming that stacking could lead to nonlinear elasticity of the PF springs and following the quantum calculations of the stacking energy for different configurations of heterocyclic rings\cite{sahu2015}, the cis-configuration of oligo-PF with heteroatoms of the $5$- and $6$-member heterocyclic rings on one side of a polymer chain was selected (see Figure \ref{fig:fig1}a)). The probing samples of the PF-springs were then preliminarily screened by molecular dynamic simulations to specify the spring sizes and the spring tensile that proved the non-linear elasticity of the spring. The distance between the adjacent turns was close to $\SI{0.35}{\nano\metre}$ in all non-stretched samples according to \citet{sahu2015}.

Guided by the preliminary screening of sizes, we designed two PF springs models consisting of five monomer units (oligo-PF-$5$) as shown in Figure \ref{fig:fig1}b,c), and seven monomer units (oligo-PF-$7$). The specificity of these models indicated that each of those springs had only one turn between the ends. It should be noted that longer PF-springs with several turns typically had many degrees of freedom associated with the movements of the turns relative to each other. These inter-turn movements made the long-term dynamics of the spring ambiguous when compared with the Duffing dynamics given by the equation \ref{eq:1}. Thus, in fact, the oligo-PF-$5$ and -$7$ springs were chosen according to the desire to have springs as short as possible, providing on the one hand, a helix-like shape of the oligomer with the stacking of aromatic groups, and on the other hand, a well-defined degree of freedom associated with long-term spring dynamics.

\subsection{Simulation details}\label{sec:MatMet3}
The oligo-PF-springs and the environmental water were modeled in a fully atomistic representation with a canonical (symbol/volume/temperature [NVT]) ensemble (box size: $\SI{7.0}{} \times \SI{7.0}{} \times \SI{7.0}{\nano\metre\cubed}$) with a time step of $\SI{2}{\femto\second}$ using Gromacs $2019$\cite{abr2015} and the OPLS-AA\cite{kam2001} force field parameters for the oligomer, and the SPC/E model\cite{ber1987} for water (for more details, see Parameters for Molecular Dynamics simulation section of Supporting Information
). The temperature was set at $\SI{280}{\kelvin}$ by the velocity-rescale thermostat\cite{bussi2007canonical}, which corresponds to the equilibrium state of PF-springs\cite{sahu2015}. Each dynamic trajectory was $\SI{300}{} - \SI{350}{\nano\second}$ long and was repeated three times to obtain better statistics; therefore. the effective length of the trajectories was about one $\SI{}{\micro\second}$ for each sample.

When studying the dynamics of the oligo-PF-$5$, one end of the spring was fixed, while the other end was pulled by a force applied along the axis of the spring. The distance (denoted  $R_{e}$) between the ends of the oligo-PF-$5$ spring (yellow and blue balls in Figure \ref{fig:fig2}a)) was considered a collective variable describing the long-term dynamics of the spring. Bistability of the oligo-PF-$5$ spring was specified in agreement with two well reproduced states of the spring with the end-to-end distances equal to $R_{e} \sim \SI{1.10}{\nano\metre}$ and  $R_{e} \sim \SI{1.45}{\nano\metre}$. These states are referred to as the squeezed and the stress-strain states, respectively.

The tensile of the oligo-PF-$7$ springs was modeled in a different way. We did not use a pulling force in this case, yet the distance between the fixed ends of the spring was the controlling parameter. Since the spring ends attracted the turn due to the $\pi-\pi$ interactions, the turn could sway between the fixed ends in a manner mimicking a pendulum. Accordingly, the states of the oligo-PF-$7$ were described by the distance $P$ between a marked atomic group on the turn and one of the spring ends (the left one). Bistability of the oligo-PF-$7$ spring was specified by means of two well reproduced positions of the turn with $P \sim \SI{0.65}{\nano\metre}$, and $P \sim \SI{0.40}{\nano\metre}$, respectively. Since these two states are associated with the closeness of the turn either to the left or right end of the spring, we refer to these states as the left- and the right-end states of the spring.

The statistics of the two states were extracted directly from the $R_{e}(t)$ and $P(t)$ series. The spectral characteristics of spontaneous vibrations and stochastic resonance were defined by the power spectra calculated using the Fourier transform of the autocorrelation functions $R_{e}(t)$ and $P(t)$, respectively.

\section{Results}
\subsection{Bistable dynamics of oligo-PF-5 spring}\label{sec:Res1}

\begin{figure}
    \centering
    \includegraphics[width=\linewidth]{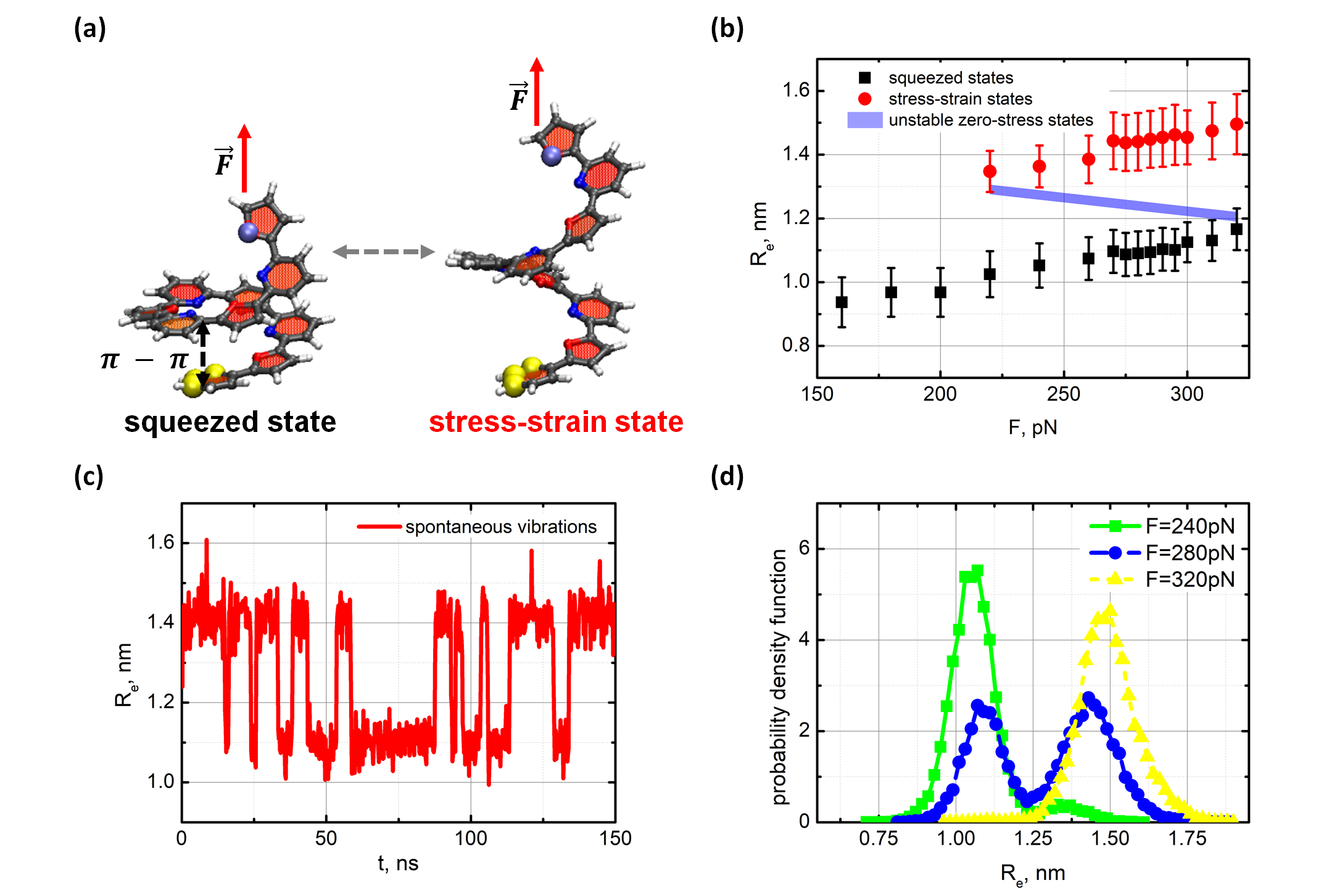}
    \caption{(a) Computational models of the oligo-PF-$5$ system with applied longitudinal load. The squeezed and the stress–strain states of the spring are shown on the left and right, respectively. The yellow spheres at the lower end of the spring indicate the fixation of the pyridine ring by rigid harmonic force. The pulling force, $F$, is applied to the top end of the spring. (b) The state diagram shows a linear elasticity of oligo-PF-$5$ spring up to  $F\approx \SI{220}{\pico\newton}$ and bistability of the spring in the region from $F\approx\SI{220}{}- \SI{320}{\pico\newton}$; (c) Spontaneous vibrations of the oligo-PF-5 spring at  $F\approx \SI{279}{\pico\newton}$; (d) Evolution of the probability density for the squeezed and stress–strain states when pulling force surpasses the critical value.}
    \label{fig:fig2}
\end{figure}

To examine the dynamics of the oligo-PF-$5$ springs that were subject to the tension, the oligo-PF-$5$ spring was first equilibrated at $\SI{280}{\kelvin}$ with one end fixed, and then pulled another end by the force $\vec{F}$ directed along the spring axis. Under a weak tensile condition, the initial state squeezed by the stacking remained stable; the spring was stretched slightly in accordance with the linear elasticity. However, as soon as the pulling force reached a specified critical value, the oligo-PF-$5$ spring became bistable and started to vibrate spontaneously. Atom level snapshots of these two states are shown in Figure \ref{fig:fig2}a). The critical value of the pulling force can be well seen in the state diagram shown in Figure \ref{fig:fig2}b). Under weak tensile conditions, only one zero-stress point can be found, which linearly shifts in accordance with the increase in pulling force (black points in Figure \ref{fig:fig2}b)). Hence, the damped oscillations characterize the spring dynamics under weak tensile conditions. As soon as the pulling force reached the critical value about of $F_{c}=\SI{240}{\pico\newton}$, a junction point occurred, which then split onto the branch of zero-stress attractors (red points on Figure \ref{fig:fig2}b)), which is referred to as a stress–strain state, and the branch of unsteady zero-stress states repulsed the dynamic trajectories (solid line in Figure \ref{fig:fig2}b)).

 At the same time, the squeezed states remain attractive (black points on Figure \ref{fig:fig2}b)). In terms of nonlinear dynamical systems, the oligo-PF-$5$ spring dynamics bifurcate at the critical force $F_{c}=\SI{240}{\pico\newton}$. Above the critical tensile, the spring becomes bistable and spontaneously vibrates between the squeezed and the stress–strain states. The mean value of the end-to-end distances of the spring in the squeezed and the stress–strain states differs by approximately $\SI{0.35}{\nano\metre}$, so the stress–strain states can be clearly distinguished from the squeezed states. Note this difference implies extending the stacking pair length to almost twice its original size. Therefore, the $\pi-\pi$ interactions do not contribute significantly to the elastic energy of the stress–strain states, and the spring elasticity is mainly determined by the rigidity of the oligomeric backbone.

Figure \ref{fig:fig2}d) shows the evolution of the statistics of visits to the squeezed and the stress–strain states when the pulling force surpasses the critical point $F_{c}$. Below $F_{c}$, the squeezed state was the only steady state of the spring. At the bifurcation point $F_{c}$, the stress–strain state appeared, and the oligo-PF-$5$ spring became bistable; it spontaneously vibrated, yet the squeezed state dominates near the critical point $F_{c}$. The squeezed and the stress–strain states were almost equally visited in the region from $F=\SI{270}{} - \SI{290}{\pico\newton}$, that is, the oligo-PF-$5$ bistability became approximately symmetrical at a point reasonably far from the critical point.

In this region, spontaneous vibrations of the oligo-PF-$5$ spring are the most pronounced. The mean lifetimes of the squeezed and the stress-strain states in the spontaneous vibrations mode varied in the bistability region from $\tau=\SI{1}{}-\SI{40}{\nano\second}$, depending on the pulling force (see Mean lifetime estimation section of Supporting Information). In the symmetrical bistability region neither the squeezed state nor the stress–strain state dominated, so the mean lifetimes of the two states were approximately the same and equal to $\tau=\SI{6.14}{\nano\second}$. In the symmetrical bistability region, spontaneous vibrations of the oligo-PF-$5$ spring were the most pronounced. Following Kramer’s rate approximation with the collision time for random perturbations ranged from $0.1-\SI{10}{\pico\second}$, one can roughly estimate the bistability barrier of the oligo-PF-$5$ spring as $10-15$ $k_{B}T$. Interesting, the bistability barrier of the oligo-PF-$5$ spring turned out to be roughly equal to the same value as that for the oligomeric Euler arch described in  \citep{avetisov2019oligomeric,markina2020detection}. Even though the reasons for bistability of the oligo-PF-$5$ spring and the oligomeric Euler arch were different, the bistability barriers of both bistable oligomeric systems were about ten times larger than the characteristic scale of thermal fluctuations, $k_{B}T$.

\begin{figure}
    \centering
    \includegraphics[width=\linewidth]{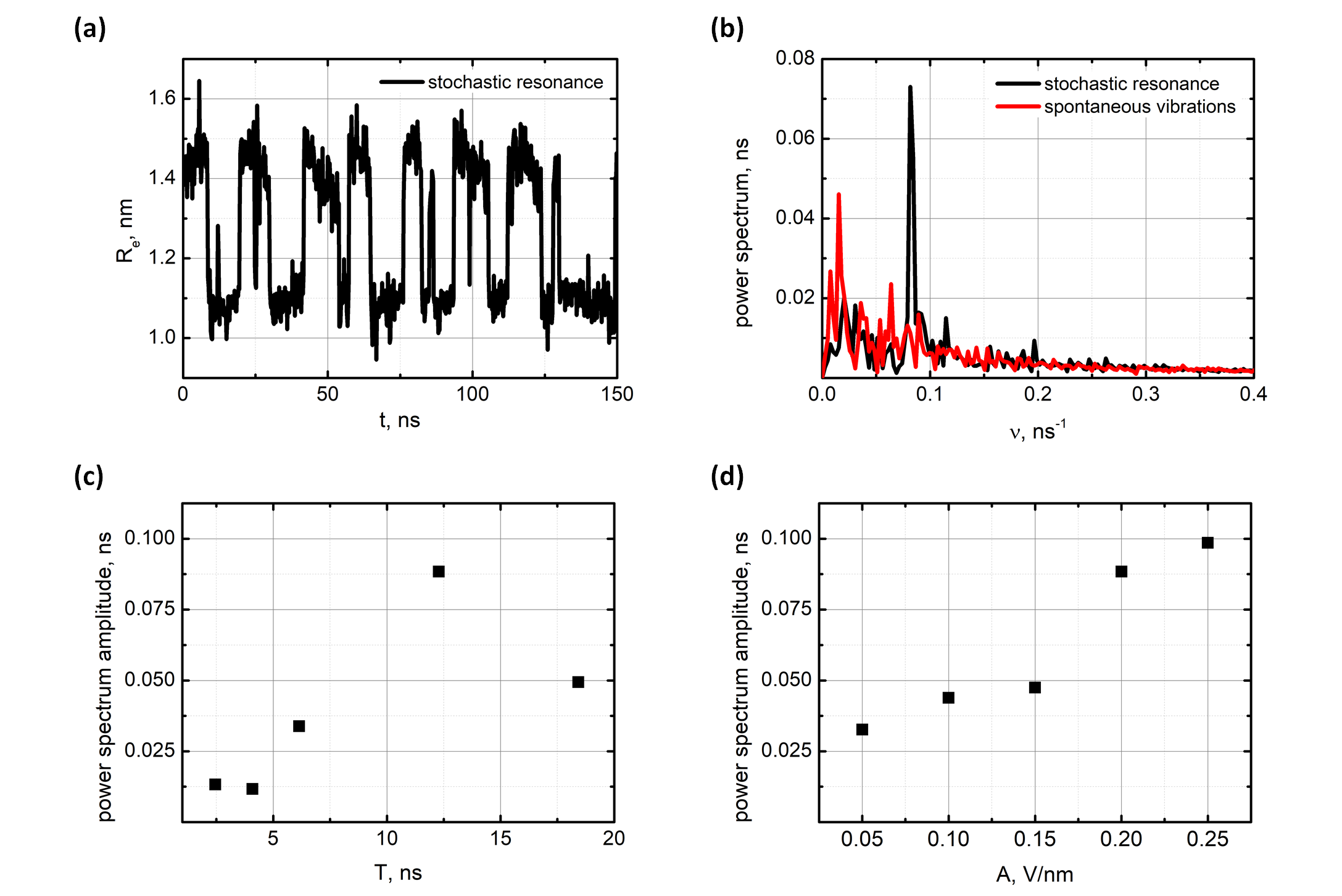}
    \caption{Stochastic resonance of the oligo-PF-5 induces by an oscillating field $E = E_{0} \cos (2 \pi \nu t)=E_{0} \cos (\nicefrac{2 \pi t}{T})$: (a) The dynamic trajectory at $F = \SI{279}{\pico\newton}$, $T = \SI{12.28}{\nano\second}$, and $E = \SI{0.2}{\volt\per\nano\metre}$; (b) Power spectrum of spontaneous vibrations (red curve) and stochastic resonance (black curve); (c) The dependence of the main resonance peak amplitude on the period $T$ of oscillating field ($E_{0}=\SI{0.2}{\volt\per\nano\metre}$); (d) The dependence of the main resonance peak amplitude on $E_{0}$ ($T=\SI{12.28}{\nano\second}$).}
    \label{fig:fig3}
\end{figure}

Figure \ref{fig:fig2}c) shows a typical trajectory of the long-term dynamics $R_{e}(t)$ of the oligo-PF-$5$ spring in the symmetric bistability region. Spontaneous vibrations of the spring could be seen unambiguously. Note, no extra random perturbations were applied to the spring to activate spontaneous vibrations as they were activated purely by the thermal-bath fluctuations. Outside the bistability region non-vibrating trajectories could be found (see Spontaneous vibrations data section of Supporting Information).

Next, we examined the stochastic resonance mode of the oligo-PF-$5$ spring by applying an additional oscillating force that waved weakly by pulling end of the spring. The oscillating force was modeled by the action of an oscillating electrical field, $E=E_{0} \cos (2\pi \nu t)$, on a unit charge preset on the pulling end of the spring while a compensative charge was on the fixed end (for more details, see Parameters of periodic signal section of Supporting Information). Typical vibrations of the end-to-end distance of the oligo-PF-$5$ spring in the stochastic resonance mode and the power spectrum of the vibrations are shown in Figure \ref{fig:fig3}a).

In accordance with the theory of stochastic resonance \cite{gammaitoni_stochastic_1998,wellens_stochastic_2004}, the main resonance peak was observed at the frequency, $\nu=\nicefrac{1}{2\tau}$, that is, the period of the applied oscillating field was equal to twice the mean lifetime of the state in the spontaneous vibration mode. In fact, we scanned a wide range of oscillating fields to find the maximal resonance response defined in terms of the spectral component at the resonance frequency. Corresponding results are presented in Figure \ref{fig:fig3}c-d). The maximum resonance response was observed exactly when the period of the oscillating field was close to twice the mean lifetime of the states in the spontaneous vibration mode. Regarding the amplitude of the oscillating field, the maximum resonance was found for $E_{0}=\SI{0.2}{\volt\per\nano\metre}$. It should be noted that the resonance response was screened in the region of symmetric bistability at $F=\SI{279}{\pico\newton}$. Beyond the symmetric bistability region, the lifetimes of the squeezed and the stress–strain states became so different that the average lifetime ceased to be a good guide for resonance frequency. 

\subsection{Bistable dynamics of oligo-PF-7 spring}\label{sec:Res2}

\begin{figure}[H]
    \centering
    \includegraphics[width=\linewidth]{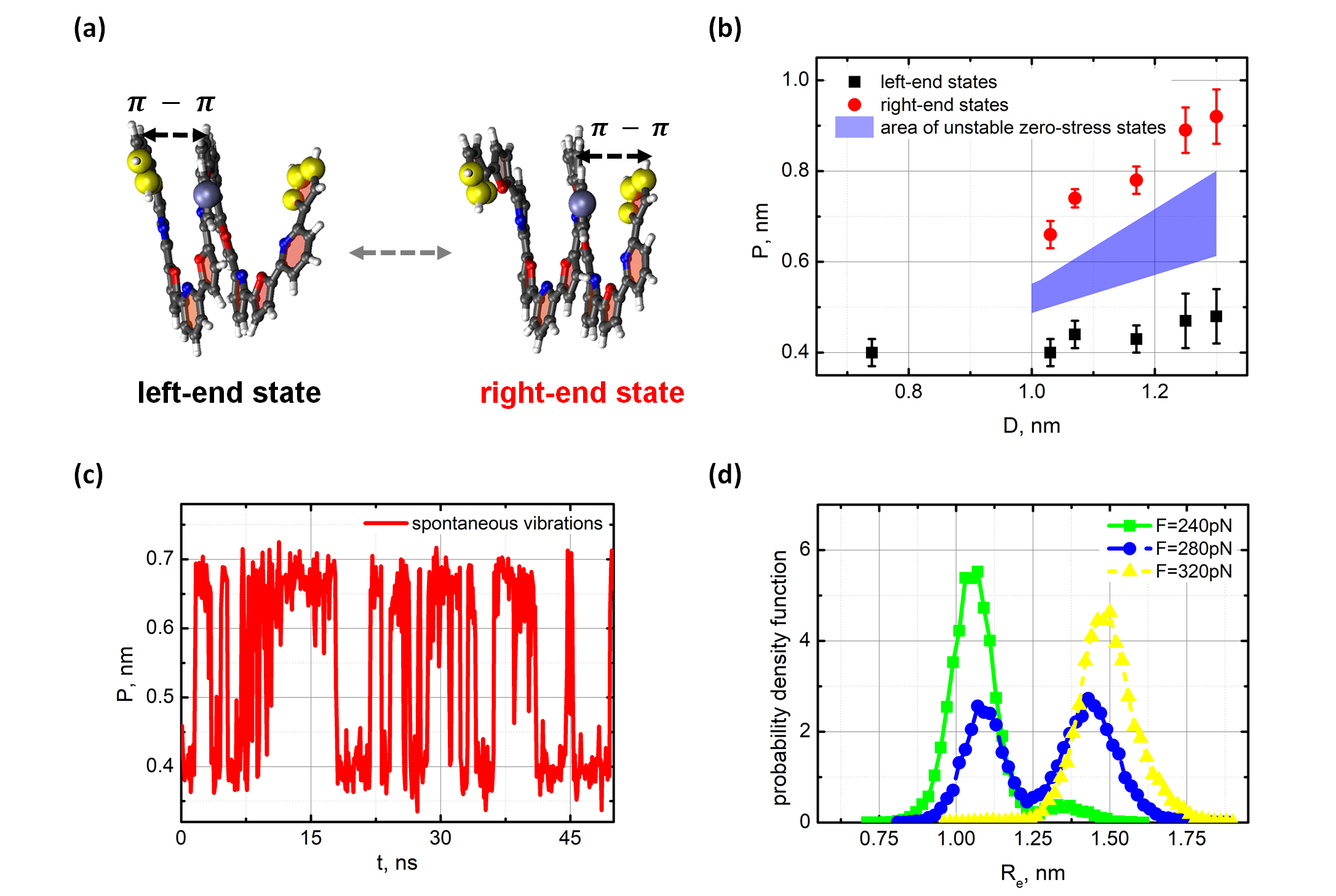}
    \caption{Bistability of the oligo-PF-$7$ spring. (a) The left- and the right-end states of the oligo-PF-$7$ (yellow spheres show fixed atomic groups at the ends of the spring); (b) State diagram of the oligo-PF-$7$ states with increasing  the end-to-end distance $D$ (c) Spontaneous vibration of the turn between the ends of the oligo-PF-$7$ spring at $D=\SI{1.03}{\nano\metre}$ ($P$ is the distance between the turn and the left end of the spring);  (d) Evolution of the probability distribution density for visiting the left-end and the right-end states at different distances $D$; almost symmetric distribution is seen at $D=\SI{1.03}{\nano\metre}$.}
    \label{fig:fig4}
\end{figure}
We also considered another aspect of the PF-springs with the Duffing bistability, which was based on mere competition between the stacking sites. The idea was to create two competing attractors at the ends of the spring so that the turn would swing between the ends like a pendulum. With that in mind, a slightly longer PF-oligomer with seven monomer units, oligo-PF-$7$ spring was designed so that the aromatic groups on the turn could form a stacking pair with either the right or left end of the spring. The equilibrated state of the oligo-PF-$7$ spring matched the squeezed state of the oligo-PF-$5$ spring with the stacking distance close to $\SI{0.35}{\nano\metre}$. In this state, the distance between the ends of the spring was about $\SI{0.7}{\nano\metre}$, and the turn was in the middle between the ends. Therefore, two possible artificial equilibrium states were created by stretching the oligo-PF-$7$ spring, thus forcing the turn to form staking pair either with the left end or with the right end of the spring. We refer these states as the left- and the right-end states, respectively. Atom level snapshots of these two states are shown in Figure \ref{fig:fig4}a).

Then, a set of oligo-PF-$7$ springs with different distances $D$ between the fixed ends was examined to search for the tensile condition resulting in bistability of the spring. The diagram of the spring states when the end-to-end distance increased is shown in Figure \ref{fig:fig4}b). Up to the distance, $D=\SI{1.0}{\nano\metre}$, the tensile was weak, and the oligo-PF-$7$ turn fluctuated somewhere around the middle. These weakly stretched oligo-PF-$7$ sspring states were the same as the squeezed states of the oligo-PF-$5$. The median zero-stress point at the middle between the spring ends was a single attractor for the spring dynamics. However, as soon as the distance $D$ exceeded $\SI{1.0}{\nano\metre}$, an extra attractive point appeared. If the spring was originally in the left-end state, the right-end states (red points in Figure \ref{fig:fig4}b)) were new attractive points, while the bistability barrier was specified by a repulsive area separating the right- and the original left-end states. The right- and left-end states were distant from each other at $\SI{0.35}{\nano\metre}$, so they were clearly distinguished. The picture was symmetrically reflected when the spring originally was in the right-end state.

Thus, the distance $D=\SI{1.0}{\nano\metre}$ between the ends corresponded to the critical tensile at which the spring states underwent bifurcation. Above the critical tensile, the oligo-PF-$7$ spring became bistable and could vibrate spontaneously between the left- and right-end states. Near the critical tensile, the left-end state of the spring was dominant if the spring originally was in this state. In symmetrical situation, the right-end site was dominant.  However, the visiting statistics of the left- and the right-end states turned out to be very sensitive with respect to the tensile of the spring. In fact, such sensitivity could be expected since the $\pi-\pi$ interactions of aromatic groups degrade sharply with stretching of a stacking pair. In our simulations, the two states of the oligo-PF-$7$ spring became almost equally visited at the distance $D=\SI{1.03}{\nano\metre}$. The spontaneous vibrations trajectory related to $D=\SI{0.03}{\nano\metre}$ is shown in Figure \ref{fig:fig4}c). The vibrations are almost symmetric, and the mean lifetimes of the left- and  right-end states were both close to $\tau=\SI{6.5}{\nano\second}$ (see Supporting Information for technical details of the mean lifetime estimations). This fact can be taken as evidence that the bistability barrier of the oligo-PF-$7$ was approximately the same as that of the oligo-PF-$5$ spring that was extended via pulling of the spring end.

\begin{figure}
    \centering
    \includegraphics[width=\linewidth]{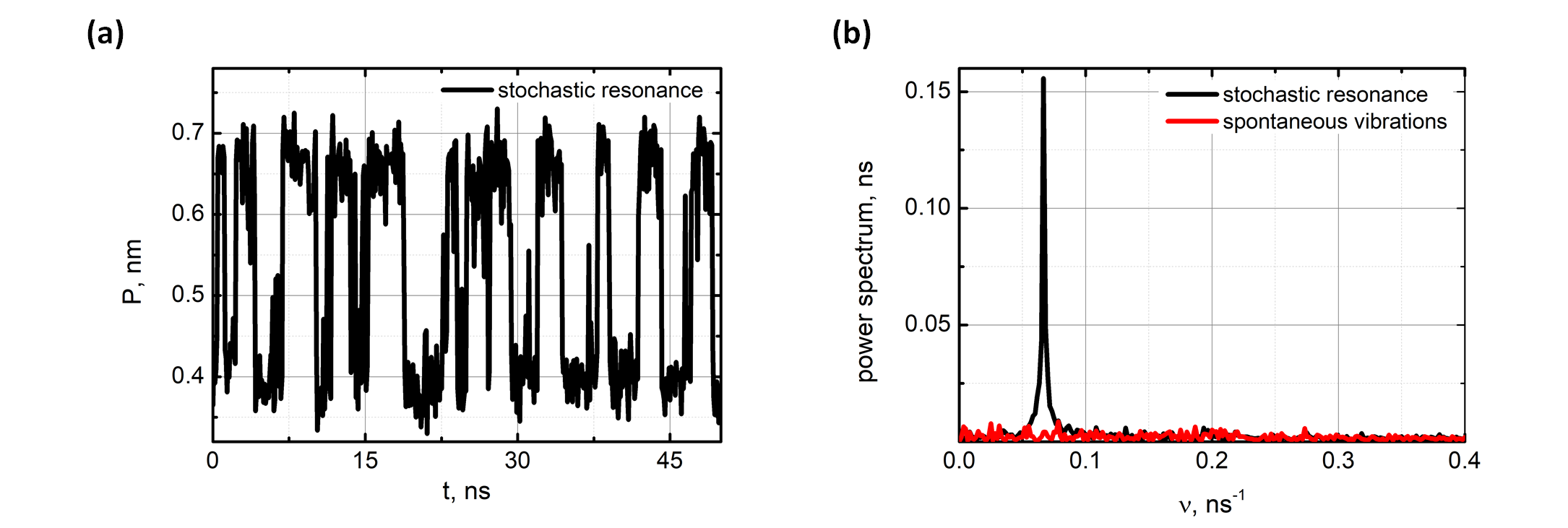}
    \caption{Stochastic resonance of the oligo-PF-$7$ induced by an oscillating field $E = E_{0} \cos (2 \pi \nu t)=E_{0} \cos (\nicefrac{2 \pi t}{T})$: (a) The dynamic trajectory at $D=\SI{1.03}{\nano\metre}, T=\SI{13}{\nano\second}$ and $E = \SI{0.2}{\volt\per\nano\metre}$. (b) Power spectrum of spontaneous vibrations (red curve) and stochastic resonance (black curve).}
    \label{fig:fig5}
\end{figure}

In addition to spontaneous vibrations, the stochastic resonance mode of the oligo-PF-$7$ spring were examined by applying a weak oscillating force to the turn of spontaneously vibrating spring. The oscillating force was implied by the action of an oscillating electrical field $E = E_{0} \cos (2 \pi \nu t)$ on a unit charge preset on the turn, while a compensating charge was put in the simulation box fettled by water molecules (for the details see Supporting Information). Typical vibrations of the oligo-PF-$7$ spring in the stochastic resonance mode are shown in Figure \ref{fig:fig5}a). The power spectrum of the vibrations is shown in Figure \ref{fig:fig5}b).

The power spectrum unambiguously highlights the stochastic resonance peak. The resonance was obtained with the oscillating field intensity of $E_{0}=\SI{0.2}{\volt\per\nano\metre}$ and frequency of $\nu \approx \nicefrac{1}{2 \tau}$, in which $\tau$ was the mean lifetime of the state in the spontaneous vibration mode. As in the case of stochastic resonance of the oligo-PF-$5$ spring, a wide range of amplitudes and frequencies of the oscillating field was scanned to find the maximal response for the stochastic resonance mode (see Signal-to-noise ratio in resonance section of Supporting Information). Guided by this scanning, the most representative conditions for the stochastic resonance were chosen, which were provided by the oscillating field with the period close to $T=2\tau=\SI{13}{\nano\second}$ and the amplitude $E_{0}=\SI{0.2}{\volt\per\nano\metre}$.

\section{Discussion}\label{Discussion}

The very idea that the competition between two attractive sites can lead to bistability is obvious. The oligo-PF-$7$ spring, the fixed ends of which play the role of two sites rivaling for the formation of stacking pairs with the spring turn, explicitly reflect this idea. However, the observation of bistability and the spontaneous vibrations in such small systems as the oligo-PF-$5$ or the oligo-PF-$7$ molecules is associated with certain obstacles. Those encountered in our simulations of spontaneous vibrations and stochastic resonance of the oligo-PF-$7$ and oligo-PF-$5$ springs are discussed above

Following the idea of two competing attractors, suppose that the pendulum-like behavior of the turn in oligo-PF-$7$ spring is controlled by two overlapping stacking potentials associated with the left and right ends of the spring. Thus, a phenomenological potential of the system can be written as described below:

\begin{align}
    \label{eq:3} U(x) = U_{stack}(x)+U_{stack}(D-x), \\
    \label{eq:4} U_{stack}(x) = A\bigg[\Big(\frac{2x_{min}} {x}\Big)^m - \Big(\frac{2x_{min}} {x}\Big)^n\bigg], 
\end{align}

in which $m>n>0$, $x_{min}$ is the stacking pair length in the ground-state associated with the minimum of a stacking potential $U_{stack}(x)$, $2x_{min}$ is the cutoff radius of the stacking, and $D$ is the distance between the fixed ends of the spring. $x_{min}$ can be set at $\SI{0.35}{\nano\metre}$ in accordance with \citeauthor{sahu2015}. After considering the motion of a particle of unit mass in the potential described by equation \ref{eq:3}, one can see that if $D$ is close to $2x_{min}$, the overlap of the stacking potentials $U_{stack}(x)$ and $U_{stack}(D-x)$ yields a degenerated minimum right in the middle of the end-to-end distance, so the particle will oscillate around $x=\nicefrac{D}{2}$. This phenomenological reasoning indicates that bistability should be expected for distances $D$ exceeding the lower limit of $d\approx 2x_{min}=\SI{0.70}{\nano\metre}$.

Formal consideration of equation \ref{eq:3} may lead to the conclusion that the potential $U(x)$ is bistable for any $D>\SI{0.70}{\nano\metre}$, so spontaneous vibrations may be expected, such as for $d\approx\SI{0.80}{\nano\metre}$. However, an addition limitation should be considered. If the bistability barrier is approximately $\leq k_B T$, the right and the left-end states of the oligo-PF-$7$ spring will then be indistinguishable against the background of fluctuations of dynamic trajectory, and the spontaneous vibrations will not be established. To observe spontaneous vibrations, the bistability barrier should be substantially greater than $k_B T$. Higher bistability barriers appear when the end-to-end distances are sufficiently longer than the lower limit of $D=\SI{0.70}{\nano\metre}$. Indeed, we observed spontaneous vibrations of the oligo-PF-$7$ spring at distances $D$ about of $3-4$ stacking lengths ranging from $\SI{1.00}{}-\SI{1.30}{\nano\metre}$.

On the other hand, if the distance between the ends of the oligo-PF-$7$ spring is larger than twice the cutoff length of stacking interactions, a wide zone of the zero-stress states will arise in the middle of the end-to-end distance in which the turn will predominantly fluctuate. This was exactly what we observed at distances $D>\SI{1.30}{\nano\metre}$ (for details, see Figure S$6$).

Thus, to observe the spontaneous vibrations of the turn between the ends, the precise adjustment of the distance between the spring's ends is obligatory. Such requirement, however, seems natural for short pyridine-furan springs since the $\pi-\pi$ interactions are short-ranged, and only one stacking pair is formed with the left or the right end of the oligo-PF-$7$ spring. Multiple stacking pairs suggest more soft control, so the requirement for the fine-tuning of the end-to-end distance might be weaker.

The next note concerns the stochastic resonance of the oligo-PF-$7$ spring. If an external oscillating field that drives the spontaneous vibrations of the turn is strong enough, the turn will subsequently move between the ends with the frequency of the oscillating field. Such forced oscillations may have nothing to do with the stochastic resonance because the stochastic resonance frequency is determined by the lifetimes of the states in the spontaneous vibration mode. Therefore, when dealing with stochastic resonance mode, the limitation on the amplitude of applied oscillating field should also be considered. Based on our simulations, we set $E_0 = \SI{0.3}{\volt\per\nano\metre}$ as the upper limit, below which the stochastic resonance was established (see Figure S$7$a).

An additional note concerns the oligo-PF-$5$ spring. In fact, bistability of PF-springs may be expected since competing interactions associated with the stacking and the backbone elasticity are found. Indeed, if the stacking interaction between the turn and the fixed end of the oligo-PF-$5$ spring controls the spring elasticity under low tensile conditions and decreases with stretching, while the elasticity imposed by oligomeric backbone stiffness increases and becomes dominant, a branch of new steady states of the spring can appear and the spring can become bistable.  What is striking is that an oligomeric molecule a few nanometers in size with only one stacking pair yields the appropriately competing interactions. Interestingly, both states of the oligo-PF-$5$ spring, the squeezed and the stress–strain states, shift with the pulling of the spring, yet the distance between these two states remains approximately the same and is equal to approximately $\SI{0.30}{\nano\metre}$. It is noteworthy that two ground states of the PF springs, which were specified using quantum calculations corresponding to “good” and “poor” accounting of the $\pi-\pi$ interactions\cite{sahu2015}, had the same difference in distances between the adjacent turns.

The region of bistability of the oligo-PF-$5$ spring was limited by the pulling force equal to approximately $F_{dest}=\SI{330}{\pico\newton}$ due to overstretching of the spring. Large pulling will irreversibly destroy the helix shape of the oligo-PF-$5$ spring, so the spring irreversibly transitions into the overstretched state after some vibrations. A greater pulling force exceeded the value $F=\SI{330}{\pico\newton}$ for which the faster transitions occurred. Once the spring reached the overstretched state, it would no longer return to the squeezed and the stress–strain state. This process was the reason that the overstretched states were not within the framework of this study.

\section{Conclusion}\label{Conclusion}

We performed the atomic level simulations of short PF-springs that were subject to stretching and found that some of the springs clearly exhibited bistable dynamics characteristic of Duffing oscillators. We studied the dynamics of two short springs designed from PF-oligomers with five and seven monomer units. When studying the dynamics of the oligo-PF-$5$ spring, one end of the spring was fixed, while another end was pulled by the force applied along the axis of the spring. The tensile of the oligo-PF-$7$ springs had been achieved by fixing both ends of the spring on controlled distance. Typical characteristics of bistability, such as spontaneous vibrations and stochastic resonance, were established for both springs and were examined in wide ranges of controlling parameters to find the symmetrical bistability conditions. At these conditions, we defined the mean lifetime of the states in the spontaneous vibration mode for each spring.  Based on these lifetimes and following Kramer’s rate approximation with the collision time ranging from $\SI{0.1}{}-\SI{10}{\pico\second}$, we estimated the bistability barriers of both springs as $10-15 k_{B}T$. It is noteworthy that the time scales of spontaneous vibrations of the oligo-PF-5 and the oligo-PF-7 springs and their bistability barriers were approximately the same as those of the oligomeric Euler arch described in  \cite{avetisov2019oligomeric,markina2020detection}. The bistability barriers of short PF springs appear to be high enough to separate the two states against the thermal noise; on the other hand, the same barriers allowed activation of the transitions between the two states by energetically enriched thermal fluctuations.

Thus, our modeling of short PF springs and antecedent modeling of the oligomeric Euler arches suggest some reasons to believe that nano-sized oligomeric structures stabilized by short-range low-energy couplings, such as by weak hydrogen bonds, hydrophilic-hydrophobic interactions, and $\pi-\pi$ interactions, can exhibit bistability with thermally-activated spontaneous vibrations and stochastic resonance. However, the proof requires challenging experimentations.

\section*{Conflicts of interest}
There are no conflicts to declare.

\section*{Data Availability Statement}
\begin{center}
\renewcommand\arraystretch{1.2}
\begin{tabular}{| >{\raggedright\arraybackslash}p{0.3\linewidth} | >{\raggedright\arraybackslash}p{0.65\linewidth} |}
\hline
\textbf{AVAILABILITY OF DATA} & \textbf{STATEMENT OF DATA AVAILABILITY}\\  
\hline
Data available on request from the authors
&
The data that support the findings of this study are available from the corresponding author upon reasonable request.
\\\hline
\end{tabular}
\end{center}
\section*{Supporting Information}
\label{SupInfo}
Supporting Information available
\bibliographystyle{unsrtnat}
\bibliography{references}

\begin{thebibliography}{51}
\providecommand{\natexlab}[1]{#1}
\providecommand{\url}[1]{\texttt{#1}}
\expandafter\ifx\csname urlstyle\endcsname\relax
  \providecommand{\doi}[1]{doi: #1}\else
  \providecommand{\doi}{doi: \begingroup \urlstyle{rm}\Url}\fi

\bibitem[Peschot et~al.(2015)Peschot, Qian, and Liu]{mi6081046}
Alexis Peschot, Chuang Qian, and Tsu-Jae~King Liu.
\newblock Nanoelectromechanical switches for low-power digital computing.
\newblock \emph{Micromachines}, 6\penalty0 (8):\penalty0 1046--1065, 2015.

\bibitem[Varghese et~al.(2015)Varghese, Elemans, Rowan, and Nolte]{C5SC02317C}
Shaji Varghese, Johannes A. A.~W. Elemans, Alan~E. Rowan, and Roeland J.~M.
  Nolte.
\newblock Molecular computing: paths to chemical turing machines.
\newblock \emph{Chem. Sci.}, 6:\penalty0 6050--6058, 2015.

\bibitem[Erbas-Cakmak et~al.(2018)Erbas-Cakmak, Kolemen, Sedgwick,
  Gunnlaugsson, James, Yoon, and Akkaya]{C7CS00491E}
Sundus Erbas-Cakmak, Safacan Kolemen, Adam~C. Sedgwick, Thorfinnur
  Gunnlaugsson, Tony~D. James, Juyoung Yoon, and Engin~U. Akkaya.
\newblock Molecular logic gates: the past{,} present and future.
\newblock \emph{Chem. Soc. Rev.}, 47:\penalty0 2228--2248, 2018.

\bibitem[Benda et~al.(2019)Benda, Doistau, Rossi-Gendron, Chamoreau,
  Hasenknopf, and Vives]{benda_substrate-dependent_2019}
Lorien Benda, Benjamin Doistau, Caroline Rossi-Gendron, Lise-Marie Chamoreau,
  Bernold Hasenknopf, and Guillaume Vives.
\newblock Substrate-dependent allosteric regulation by switchable catalytic
  molecular tweezers.
\newblock \emph{Communications Chemistry}, 2\penalty0 (1):\penalty0 1--11,
  2019.

\bibitem[Berselli et~al.(2021)Berselli, Gimenez, O’Connor, and
  Keyes]{berselli_robust_2021}
Guilherme~B. Berselli, Aurélien~V. Gimenez, Alexandra O’Connor, and Tia~E.
  Keyes.
\newblock Robust {Photoelectric} {Biomolecular} {Switch} at a
  {Microcavity}-{Supported} {Lipid} {Bilayer}.
\newblock \emph{ACS Applied Materials \& Interfaces}, 13\penalty0
  (24):\penalty0 29158--29169, 2021.

\bibitem[Nicoli et~al.(2021)Nicoli, Paltrinieri, {Tranfić Bakić}, Baroncini,
  Silvi, and Credi]{NICOLI2021213589}
Federico Nicoli, Erica Paltrinieri, Marina {Tranfić Bakić}, Massimo
  Baroncini, Serena Silvi, and Alberto Credi.
\newblock Binary logic operations with artificial molecular machines.
\newblock \emph{Coordination Chemistry Reviews}, 428:\penalty0 213589, 2021.

\bibitem[Zhang et~al.(2018)Zhang, Marcos, and Leigh]{zhang_molecular_2018}
Liang Zhang, Vanesa Marcos, and David~A. Leigh.
\newblock Molecular machines with bio-inspired mechanisms.
\newblock \emph{Proceedings of the National Academy of Sciences}, 115\penalty0
  (38):\penalty0 9397--9404, 2018.

\bibitem[Shu et~al.(2020)Shu, Shen, Zhang, and
  Serpe]{shu_stimuli-responsive_2020}
Tong Shu, Qiming Shen, Xueji Zhang, and Michael~J. Serpe.
\newblock Stimuli-responsive polymer/nanomaterial hybrids for sensing
  applications.
\newblock \emph{Analyst}, 145\penalty0 (17):\penalty0 5713--5724, 2020.

\bibitem[Lemme et~al.(2020)Lemme, Wagner, Lee, Fan, Verbiest, Wittmann, Lukas,
  Dolleman, Niklaus, van~der Zant, Duesberg, and
  Steeneken]{lemme_nanoelectromechanical_2020}
Max~C. Lemme, Stefan Wagner, Kangho Lee, Xuge Fan, Gerard~J. Verbiest,
  Sebastian Wittmann, Sebastian Lukas, Robin~J. Dolleman, Frank Niklaus, Herre
  S.~J. van~der Zant, Georg~S. Duesberg, and Peter~G. Steeneken.
\newblock Nanoelectromechanical {Sensors} {Based} on {Suspended} {2D}
  {Materials}.
\newblock \emph{Research}, 2020, 2020.

\bibitem[Shi et~al.(2020)Shi, Zhang, Tian, and Qu]{shi_driving_2020}
Zhao-Tao Shi, Qi~Zhang, He~Tian, and Da-Hui Qu.
\newblock Driving {Smart} {Molecular} {Systems} by {Artificial} {Molecular}
  {Machines}.
\newblock \emph{Advanced Intelligent Systems}, 2\penalty0 (5):\penalty0
  1900169, 2020.

\bibitem[Aprahamian(2020)]{aprahamian_future_2020}
Ivan Aprahamian.
\newblock The {Future} of {Molecular} {Machines}.
\newblock \emph{ACS Central Science}, 6\penalty0 (3):\penalty0 347--358, 2020.

\bibitem[Li et~al.(2014)Li, Tian, and Deng]{li_energy_2014}
Huidong Li, Chuan Tian, and Z.~Daniel Deng.
\newblock Energy harvesting from low frequency applications using piezoelectric
  materials.
\newblock \emph{Applied Physics Reviews}, 1\penalty0 (4):\penalty0 041301,
  2014.

\bibitem[Kim et~al.(2015)Kim, Lima, Kozlov, Haines, Spinks, Aziz, Choi, Sim,
  Wang, Lu, Qian, Madden, Baughman, and Kim]{kim_harvesting_2015}
Shi~Hyeong Kim, Márcio~D. Lima, Mikhail~E. Kozlov, Carter~S. Haines,
  Geoffrey~M. Spinks, Shazed Aziz, Changsoon Choi, Hyeon~Jun Sim, Xuemin Wang,
  Hongbing Lu, Dong Qian, John D.~W. Madden, Ray~H. Baughman, and Seon~Jeong
  Kim.
\newblock Harvesting temperature fluctuations as electrical energy using
  torsional and tensile polymer muscles.
\newblock \emph{Energy \& Environmental Science}, 8\penalty0 (11):\penalty0
  3336--3344, 2015.

\bibitem[Ackerman et~al.(2016)Ackerman, Kumar, Neek-Amal, Thibado, Peeters, and
  Singh]{ackerman_anomalous_2016}
M.~L. Ackerman, P.~Kumar, M.~Neek-Amal, P.~M. Thibado, F.~M. Peeters, and
  Surendra Singh.
\newblock Anomalous {Dynamical} {Behavior} of {Freestanding} {Graphene}
  {Membranes}.
\newblock \emph{Physical Review Letters}, 117\penalty0 (12):\penalty0 126801,
  2016.

\bibitem[Dutreix et~al.(2020)Dutreix, Avriller, Lounis, and
  Pistolesi]{dutreix_two-level_2020}
C.~Dutreix, R.~Avriller, B.~Lounis, and F.~Pistolesi.
\newblock Two-level system as topological actuator for nanomechanical modes.
\newblock \emph{Physical Review Research}, 2\penalty0 (2):\penalty0 023268,
  2020.

\bibitem[Thibado et~al.(2020)Thibado, Kumar, Singh, Ruiz-Garcia, Lasanta, and
  Bonilla]{thibado_fluctuation-induced_2020}
P.~M. Thibado, P.~Kumar, Surendra Singh, M.~Ruiz-Garcia, A.~Lasanta, and L.~L.
  Bonilla.
\newblock Fluctuation-induced current from freestanding graphene.
\newblock \emph{Physical Review E}, 102\penalty0 (4):\penalty0 042101, 2020.

\bibitem[Evans and Searles(2002)]{evans_fluctuation_2002}
Denis~J. Evans and Debra~J. Searles.
\newblock The {Fluctuation} {Theorem}.
\newblock \emph{Advances in Physics}, 51\penalty0 (7):\penalty0 1529--1585,
  2002.

\bibitem[Seifert(2012)]{seifert_stochastic_2012}
Udo Seifert.
\newblock Stochastic thermodynamics, fluctuation theorems and molecular
  machines.
\newblock \emph{Reports on Progress in Physics}, 75\penalty0 (12):\penalty0
  126001, 2012.

\bibitem[Horowitz and Gingrich(2020)]{horowitz_thermodynamic_2020}
Jordan~M. Horowitz and Todd~R. Gingrich.
\newblock Thermodynamic uncertainty relations constrain non-equilibrium
  fluctuations.
\newblock \emph{Nature Physics}, 16\penalty0 (1):\penalty0 15--20, 2020.

\bibitem[Ciliberto(2017)]{ciliberto_experiments_2017}
S.~Ciliberto.
\newblock Experiments in {Stochastic} {Thermodynamics}: {Short} {History} and
  {Perspectives}.
\newblock \emph{Physical Review X}, 7:\penalty0 021051, 2017.

\bibitem[Wang et~al.(2002)Wang, Sevick, Mittag, Searles, and
  Evans]{wang_experimental_2002}
G.~M. Wang, E.~M. Sevick, Emil Mittag, Debra~J. Searles, and Denis~J. Evans.
\newblock Experimental {Demonstration} of {Violations} of the {Second} {Law} of
  {Thermodynamics} for {Small} {Systems} and {Short} {Time} {Scales}.
\newblock \emph{Physical Review Letters}, 89\penalty0 (5):\penalty0 050601,
  2002.

\bibitem[Jop et~al.(2008)Jop, Petrosyan, and Ciliberto]{jop_work_2008}
P.~Jop, A.~Petrosyan, and S.~Ciliberto.
\newblock Work and dissipation fluctuations near the stochastic resonance of a
  colloidal particle.
\newblock \emph{EPL (Europhysics Letters)}, 81\penalty0 (5):\penalty0 50005,
  2008.

\bibitem[Astumian(2018)]{astumian_stochastic_2018}
R.~D. Astumian.
\newblock Stochastic pumping of non-equilibrium steady-states: how molecules
  adapt to a fluctuating environment.
\newblock \emph{Chemical Communications}, 54\penalty0 (5):\penalty0 427--444,
  2018.

\bibitem[Vroylandt et~al.(2020)Vroylandt, Esposito, and
  Verley]{vroylandt_efficiency_2020}
Hadrien Vroylandt, Massimiliano Esposito, and Gatien Verley.
\newblock Efficiency {Fluctuations} of {Stochastic} {Machines} {Undergoing} a
  {Phase} {Transition}.
\newblock \emph{Physical Review Letters}, 124\penalty0 (25):\penalty0 250603,
  2020.

\bibitem[Arnold(1984)]{arnold_catastrophe_1984}
Vladimir~Igorevich Arnold.
\newblock \emph{Catastrophe {Theory}}.
\newblock Springer Berlin Heidelberg, Berlin, Heidelberg, 1984.
\newblock ISBN 9783540128595 9783642967993.

\bibitem[Poston and Stewart(1996)]{poston_catastrophe_1996}
T.~Poston and Ian Stewart.
\newblock \emph{Catastrophe theory and its applications}.
\newblock Dover Publications, Mineola, N.Y, 1996.
\newblock ISBN 9780486692715.

\bibitem[Duffing(1918)]{duffing1918erzwungene}
Georg Duffing.
\newblock \emph{Erzwungene schwingungen bei ver{\"a}nderlicher eigenfrequenz
  und ihre technische bedeutung}.
\newblock Number 41--42. F. Vieweg \& sohn, 1918.

\bibitem[kor(2008)]{korsch_duffing_2008}
The {Duffing} {Oscillator}.
\newblock In Hans~Jürgen Korsch, Hans-Jörg Jodl, and Timo Hartmann, editors,
  \emph{Chaos: {A} {Program} {Collection} for the {PC}}, pages 157--184.
  Springer, Berlin, Heidelberg, 2008.
\newblock ISBN 9783540748670.

\bibitem[Kramers(1940)]{KRAMERS1940284}
H.A. Kramers.
\newblock Brownian motion in a field of force and the diffusion model of
  chemical reactions.
\newblock \emph{Physica}, 7\penalty0 (4):\penalty0 284--304, 1940.

\bibitem[Benzi et~al.(1981)Benzi, Sutera, and Vulpiani]{benzi_mechanism_1981}
R~Benzi, A~Sutera, and A~Vulpiani.
\newblock The mechanism of stochastic resonance.
\newblock \emph{Journal of Physics A: Mathematical and General}, 14\penalty0
  (11):\penalty0 L453--L457, 1981.

\bibitem[Benzi et~al.(1982)Benzi, Parisi, Sutera, and
  Vulpiani]{benzi_stochastic_1982}
Roberto Benzi, Giorgio Parisi, Alfonso Sutera, and Angelo Vulpiani.
\newblock Stochastic resonance in climatic change.
\newblock \emph{Tellus}, 34\penalty0 (1):\penalty0 10--15, 1982.

\bibitem[Benzi et~al.(1983)Benzi, Parisi, Sutera, and
  Vulpiani]{benzi_theory_1983}
Roberto Benzi, Giorgio Parisi, Alfonso Sutera, and Angelo Vulpiani.
\newblock A {Theory} of {Stochastic} {Resonance} in {Climatic} {Change}.
\newblock \emph{SIAM Journal on Applied Mathematics}, 43\penalty0 (3):\penalty0
  565--578, 1983.

\bibitem[Gammaitoni et~al.(1998)Gammaitoni, Hänggi, Jung, and
  Marchesoni]{gammaitoni_stochastic_1998}
Luca Gammaitoni, Peter Hänggi, Peter Jung, and Fabio Marchesoni.
\newblock Stochastic resonance.
\newblock \emph{Reviews of Modern Physics}, 70\penalty0 (1):\penalty0 223--287,
  1998.

\bibitem[Wellens et~al.(2004)Wellens, Shatokhin, and
  Buchleitner]{wellens_stochastic_2004}
Thomas Wellens, Vyacheslav Shatokhin, and Andreas Buchleitner.
\newblock Stochastic resonance.
\newblock \emph{Reports on Progress in Physics}, 67\penalty0 (1):\penalty0
  45--105, 2004.

\bibitem[Baughman et~al.(1999)Baughman, Cui, Zakhidov, Iqbal, Barisci, Spinks,
  Wallace, Mazzoldi, De~Rossi, Rinzler, et~al.]{baughman1999carbon}
Ray~H Baughman, Changxing Cui, Anvar~A Zakhidov, Zafar Iqbal, Joseph~N Barisci,
  Geoff~M Spinks, Gordon~G Wallace, Alberto Mazzoldi, Danilo De~Rossi, Andrew~G
  Rinzler, et~al.
\newblock Carbon nanotube actuators.
\newblock \emph{Science}, 284\penalty0 (5418):\penalty0 1340--1344, 1999.

\bibitem[Fujii et~al.(2017)Fujii, Setiadi, Kuwahara, and
  Akai-Kasaya]{fujii2017single}
Hayato Fujii, Agung Setiadi, Yuji Kuwahara, and Megumi Akai-Kasaya.
\newblock Single walled carbon nanotube-based stochastic resonance device with
  molecular self-noise source.
\newblock \emph{Applied Physics Letters}, 111\penalty0 (13):\penalty0 133501,
  2017.

\bibitem[Huang et~al.(2019)Huang, Zhang, Li, and Li]{huang_nonlocal_2019}
Kun Huang, Shuzhu Zhang, Jinhai Li, and Ze~Li.
\newblock Nonlocal nonlinear model of {Bernoulli}–{Euler} nanobeam with small
  initial curvature and its application to single-walled carbon nanotubes.
\newblock \emph{Microsystem Technologies}, 25\penalty0 (11):\penalty0
  4303--4310, 2019.

\bibitem[Liang et~al.(2012)Liang, Huang, Li, Huang, Wu, Fang, Oh, Kozlov, Ma,
  Li, et~al.]{liang2012electromechanical}
Jiajie Liang, Lu~Huang, Na~Li, Yi~Huang, Yingpeng Wu, Shaoli Fang, Jiyoung Oh,
  Mikhail Kozlov, Yanfeng Ma, Feifei Li, et~al.
\newblock Electromechanical actuator with controllable motion, fast response
  rate, and high-frequency resonance based on graphene and polydiacetylene.
\newblock \emph{ACS nano}, 6\penalty0 (5):\penalty0 4508--4519, 2012.

\bibitem[Forns et~al.(2011)Forns, de~Lorenzo, Manosas, Hayashi, Huguet, and
  Ritort]{forns2011improving}
Nuria Forns, Sara de~Lorenzo, Maria Manosas, Kumiko Hayashi, Josep~Maria
  Huguet, and Felix Ritort.
\newblock Improving signal/noise resolution in single-molecule experiments
  using molecular constructs with short handles.
\newblock \emph{Biophysical Journal}, 100\penalty0 (7):\penalty0 1765--1774,
  2011.

\bibitem[Hayashi et~al.(2012)Hayashi, de~Lorenzo, Manosas, Huguet, and
  Ritort]{hayashi2012single}
K~Hayashi, S~de~Lorenzo, M~Manosas, JM~Huguet, and F~Ritort.
\newblock Single-molecule stochastic resonance.
\newblock \emph{Physical Review X}, 2\penalty0 (3):\penalty0 031012, 2012.

\bibitem[Cecconi et~al.(2005)Cecconi, Shank, Bustamante, and
  Marqusee]{cecconi2005direct}
Ciro Cecconi, Elizabeth~A Shank, Carlos Bustamante, and Susan Marqusee.
\newblock Direct observation of the three-state folding of a single protein
  molecule.
\newblock \emph{Science}, 309\penalty0 (5743):\penalty0 2057--2060, 2005.

\bibitem[Avetisov et~al.(2019)Avetisov, Markina, and
  Valov]{avetisov2019oligomeric}
Vladik~A Avetisov, Anastasia~A Markina, and Alexander~F Valov.
\newblock Oligomeric “catastrophe machines” with thermally activated
  bistability and stochastic resonance.
\newblock \emph{The Journal of Physical Chemistry Letters}, 10\penalty0
  (17):\penalty0 5189--5192, 2019.

\bibitem[Markina et~al.(2020)Markina, Muratov, Petrovskyy, and
  Avetisov]{markina2020detection}
Anastasia Markina, Alexander Muratov, Vladislav Petrovskyy, and Vladik
  Avetisov.
\newblock Detection of single molecules using stochastic resonance of bistable
  oligomers.
\newblock \emph{Nanomaterials}, 10\penalty0 (12):\penalty0 2519, 2020.

\bibitem[Lai and Leng(2016)]{LAI201660}
{Zhi-hui} Lai and {Yong-gang} Leng.
\newblock Weak-signal detection based on the stochastic resonance of bistable
  duffing oscillator and its application in incipient fault diagnosis.
\newblock \emph{Mechanical Systems and Signal Processing}, 81:\penalty0 60--74,
  2016.

\bibitem[Lu et~al.(2021)Lu, Wu, Ding, and Chen]{LU2021249}
Ze-Qi Lu, Dao Wu, Hu~Ding, and Li-Qun Chen.
\newblock Vibration isolation and energy harvesting integrated in a stewart
  platform with high static and low dynamic stiffness.
\newblock \emph{Applied Mathematical Modelling}, 89:\penalty0 249--267, 2021.

\bibitem[Jones and Civcir(1997)]{jones1997extended}
R~Alan Jones and Pervin~U Civcir.
\newblock Extended heterocyclic systems 2. the synthesis and characterisation
  of (2-furyl) pyridines,(2-thienyl) pyridines, and furan-pyridine and
  thiophene-pyridine oligomers.
\newblock \emph{Tetrahedron}, 53\penalty0 (34):\penalty0 11529--11540, 1997.

\bibitem[Harikrishna~Sahu and Panda(2015)]{sahu2015}
Priyank~Gaur Harikrishna~Sahu, Shashwat~Gupta and Aditya~N. Panda.
\newblock Structure and optoelectronic properties of helical pyridine–furan,
  pyridine–pyrrole and pyridine–thiophene oligomers.
\newblock \emph{Physical Chemistry Chemical Physics}, 17\penalty0
  (32):\penalty0 20647--20657, 2015.

\bibitem[Abraham et~al.(2015)Abraham, Murtola, Schulz, P{\'a}ll, Smith, Hess,
  and Lindahl]{abr2015}
Mark~James Abraham, Teemu Murtola, Roland Schulz, Szil{\'a}rd P{\'a}ll,
  Jeremy~C Smith, Berk Hess, and Erik Lindahl.
\newblock Gromacs: High performance molecular simulations through multi-level
  parallelism from laptops to supercomputers.
\newblock \emph{SoftwareX}, 1:\penalty0 19--25, 2015.

\bibitem[Kaminski et~al.(2001)Kaminski, Friesner, Tirado-Rives, and
  Jorgensen]{kam2001}
George~A Kaminski, Richard~A Friesner, Julian Tirado-Rives, and William~L
  Jorgensen.
\newblock Evaluation and reparametrization of the opls-aa force field for
  proteins via comparison with accurate quantum chemical calculations on
  peptides.
\newblock \emph{The Journal of Physical Chemistry B}, 105\penalty0
  (28):\penalty0 6474--6487, 2001.

\bibitem[Berendsen et~al.(1987)Berendsen, Grigera, and Straatsma]{ber1987}
HJC Berendsen, JR~Grigera, and TP~Straatsma.
\newblock The missing term in effective pair potentials.
\newblock \emph{Journal of Physical Chemistry}, 91\penalty0 (24):\penalty0
  6269--6271, 1987.

\bibitem[Bussi et~al.(2007)Bussi, Donadio, and Parrinello]{bussi2007canonical}
Giovanni Bussi, Davide Donadio, and Michele Parrinello.
\newblock Canonical sampling through velocity rescaling.
\newblock \emph{The Journal of Chemical Physics}, 126\penalty0 (1):\penalty0
  014101, 2007.

\end{thebibliography}


\providecommand{\latin}[1]{#1}
\makeatletter
\providecommand{\doi}
  {\begingroup\let\do\@makeother\dospecials
  \catcode`\{=1 \catcode`\}=2 \doi@aux}
\providecommand{\doi@aux}[1]{\endgroup\texttt{#1}}
\makeatother
\providecommand*\mcitethebibliography{\thebibliography}
\csname @ifundefined\endcsname{endmcitethebibliography}
  {\let\endmcitethebibliography\endthebibliography}{}
\begin{mcitethebibliography}{6}
\providecommand*\natexlab[1]{#1}
\providecommand*\mciteSetBstSublistMode[1]{}
\providecommand*\mciteSetBstMaxWidthForm[2]{}
\providecommand*\mciteBstWouldAddEndPuncttrue
  {\def\EndOfBibitem{\unskip.}}
\providecommand*\mciteBstWouldAddEndPunctfalse
  {\let\EndOfBibitem\relax}
\providecommand*\mciteSetBstMidEndSepPunct[3]{}
\providecommand*\mciteSetBstSublistLabelBeginEnd[3]{}
\providecommand*\EndOfBibitem{}
\mciteSetBstSublistMode{f}
\mciteSetBstMaxWidthForm{subitem}{(\alph{mcitesubitemcount})}
\mciteSetBstSublistLabelBeginEnd
  {\mcitemaxwidthsubitemform\space}
  {\relax}
  {\relax}

\bibitem[Abraham \latin{et~al.}(2015)Abraham, Murtola, Schulz, P{\'a}ll, Smith,
  Hess, and Lindahl]{abr2015}
Abraham,~M.~J.; Murtola,~T.; Schulz,~R.; P{\'a}ll,~S.; Smith,~J.~C.; Hess,~B.;
  Lindahl,~E. GROMACS: High performance molecular simulations through
  multi-level parallelism from laptops to supercomputers. \emph{SoftwareX}
  \textbf{2015}, \emph{1}, 19--25\relax
\mciteBstWouldAddEndPuncttrue
\mciteSetBstMidEndSepPunct{\mcitedefaultmidpunct}
{\mcitedefaultendpunct}{\mcitedefaultseppunct}\relax
\EndOfBibitem
\bibitem[Kaminski \latin{et~al.}(2001)Kaminski, Friesner, Tirado-Rives, and
  Jorgensen]{kam2001}
Kaminski,~G.~A.; Friesner,~R.~A.; Tirado-Rives,~J.; Jorgensen,~W.~L. Evaluation
  and reparametrization of the OPLS-AA force field for proteins via comparison
  with accurate quantum chemical calculations on peptides. \emph{The Journal of
  Physical Chemistry B} \textbf{2001}, \emph{105}, 6474--6487\relax
\mciteBstWouldAddEndPuncttrue
\mciteSetBstMidEndSepPunct{\mcitedefaultmidpunct}
{\mcitedefaultendpunct}{\mcitedefaultseppunct}\relax
\EndOfBibitem
\bibitem[Berendsen \latin{et~al.}(1987)Berendsen, Grigera, and
  Straatsma]{ber1987}
Berendsen,~H.; Grigera,~J.; Straatsma,~T. The missing term in effective pair
  potentials. \emph{Journal of Physical Chemistry} \textbf{1987}, \emph{91},
  6269--6271\relax
\mciteBstWouldAddEndPuncttrue
\mciteSetBstMidEndSepPunct{\mcitedefaultmidpunct}
{\mcitedefaultendpunct}{\mcitedefaultseppunct}\relax
\EndOfBibitem
\bibitem[Essmann \latin{et~al.}(1995)Essmann, Perera, and Berkowitz]{ess1995}
Essmann,~U.; Perera,~L.; Berkowitz,~M.~L. A smooth particle mesh Ewald method.
  \emph{The Journal of chemical physics} \textbf{1995}, \emph{103},
  8577--8593\relax
\mciteBstWouldAddEndPuncttrue
\mciteSetBstMidEndSepPunct{\mcitedefaultmidpunct}
{\mcitedefaultendpunct}{\mcitedefaultseppunct}\relax
\EndOfBibitem
\bibitem[Berk(2008)]{hes2008}
Berk,~H. P-LINCS: A parallel linear constraint solver for molecular simulation.
  \emph{Journal of Chemical Theory and Computation} \textbf{2008}, \emph{4},
  116--122\relax
\mciteBstWouldAddEndPuncttrue
\mciteSetBstMidEndSepPunct{\mcitedefaultmidpunct}
{\mcitedefaultendpunct}{\mcitedefaultseppunct}\relax
\EndOfBibitem
\end{mcitethebibliography}

\end{document}


\section{Simulation protocol}
\subsection{Parameters for Molecular Dynamics simulation}\label{SI1}
Morphology simulations were performed using the Gromacs2019 \cite{abr2015} simulation package. Lennard– Jones parameters were taken from the OPLS-AA\cite{kam2001} force field with a scaling factor of $0.5$ for the $1–4$ interactions. The SPC/E model \cite{ber1987} was used for water. Long-range electrostatic interactions were treated using a smooth particle mesh Ewald technique\cite{ess1995} with a cut-off of \SI{1.2}{\nano\metre}. Bond vibrations were constrained with a LINCS \cite{hes2008} algorithm. All calculations were performed in the NVT ensemble using the canonical velocity-rescaling thermostat, as implemented in the Gromacs2019 simulation package.

\begin{figure}[H]
    \includegraphics[width=\linewidth]{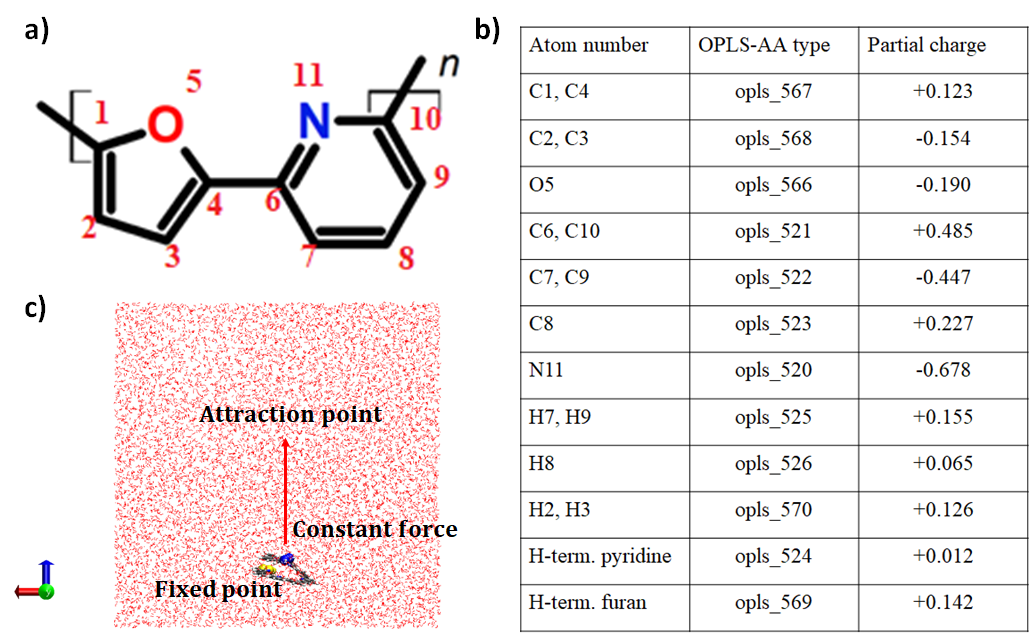}
    \centering
\caption{(a) The chemical structure of a PF monomer, (b)parameterization of PF oligomers in OPLS-AA force field types and partial charges, (c) a scheme of the simulation box for the oligo-PF5 system.}
\label{fig:S1}
\end{figure}

The parameterization of oligo-PF springs is presented in Figure \ref{fig:S1}a-b. The OPLS-AA force field contains parameters only for single pyridine and furan molecules. First, $C4$ and $C6$ atoms have a covalent bond instead of hydrogen atoms and the partial charge of hydrogens were added to carbon atoms to maintain the neutrality. Second, there is only one undefined angle bond for atoms $C4-C6-N11$, which is $CW-CA-NC$ angle type in OPLS-AA. We used $CA-CA-NC$ angle due to similarity of CA and CW types. Both correspond to aromatic $sp2$-hybridized carbon.

The simulation was started from a random initial configuration. An equilibrated state were reached in two steps. First, a short run of \SI{10}{\pico\second} was carried out in the NVT ensemble with a time step of \SI{0.01}{\femto\second} with the leapfrog integrator for motion equations. Then the second part of equilibration was done for another \SI{10}{\nano\second} with a time step of \SI{2}{\femto\second}. After the equilibration, a long trajectory of \SI{500}{\nano\second} is obtained to achieve a clear picture of transitional behaviour.  

To simulate the behaviour of the oligo-PF$5$ under the external load, one end of it was fixed and the other end was pulled by an external force (see Figure \ref{fig:S1}c). We used a simulation box of size $7\times7\times7$ $nm^3$ due to technical aspects of Gromacs pulling algorithm. The axis of an initial conformation of oligo-PF$5$ was oriented in the XY plane. The longitudinal load F was applied along Z-direction to the center of mass of the last monomer unit and directed toward the attraction point, which was located along the vector connecting the left and right ends of the molecule.

At the same time, to simulate behaviour of the oligo-PF$7$, both ends of the molecule in the stressed state were fixed in a simulation box of size $3.7\times3.7\times3.7$ $nm^3$. The different distances $P$ were obtained by pulling one end of the oligo-PF$7$. Then the pulled end was fixed and the vibrations were explored with different distances $P$. Various initial structures are shown in Figure S2.

In both cases, the center of mass of the first monomeric unit was fixed using a spring potential of $k=\SI{100}{\kilo\joule\per\mole\per\nano\metre\squared}$; other specific constraints for bond length or atom positions were applied.
 
\begin{figure}[H]
    \includegraphics[width=\linewidth]{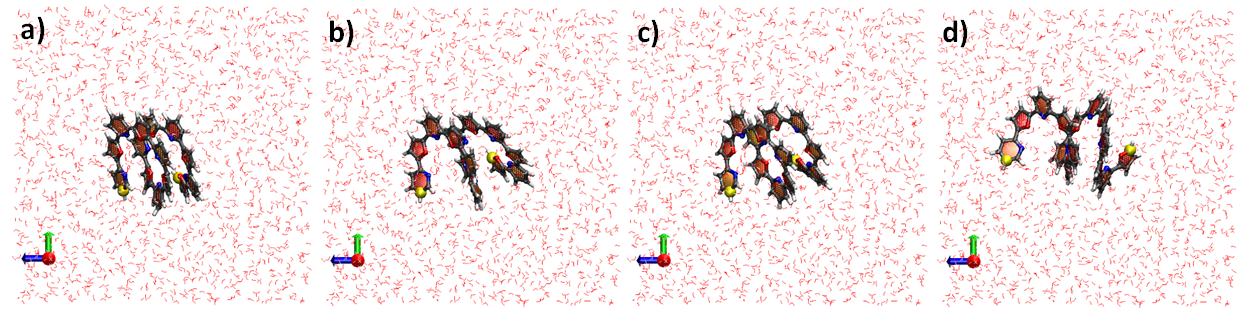}
    \centering
\caption{Initial states of the oligo-PF7 are shown with different distances $P$ = 0.74, 1.07, 1.03, and 1.62 nm between fixed points (yellow dots).}
\label{fig:S2}
\end{figure}
\subsection{Parameters for periodic signal}\label{SI:2}
Stochastic resonance was obtained by applying a periodic signal leading to the swinging of the oligo-PF bistable potential. An oscillating force was implemented by setting a charge ($+1$) on the oligomer, adding a compensating charge ($–1$) as a counter ion in the solvent and applying a periodical electric field. The additional charge was placed on the end group of the oligo-PF$5$ and spread between the first, second and fifth atoms of the furan group (see Fig. \ref{fig:S1}a). In case of the oligo-PF$7$, the charge is set on the middle fourth monomer unit and spread between the seventh, eighth and ninth atoms of the pyrydine group. In all cases the additional chargers are equally distributed between the chosen atoms.

The periodic field in the Gromacs2019 package is defined by an equation $$ E(t) = E_0 \exp{\left[-\frac{(t-t_0)^2}{2\sigma^2}\right]}\cos{\omega(t-t_0)},$$ where the exponential part modulates the periodic part with the pulsing behaviour, $E_0$ is the amplitude of the signal and $\omega t_0$ is the oscillation phase. Here we use only the static part when $\sigma=0$ along z-axes (see Figure \ref{fig:S1}c and Figure \ref{fig:S2}). An external oscillating electrical field was directed along the constant force direction in the oligo-PF$5$ case. In the oligo-PF$7$ system, the periodic signal was applied along a line connecting fixed atoms.

\section{Results}
\subsection{Spontaneous vibrations data}
\begin{figure}[H]
    \includegraphics[width=0.7\linewidth]{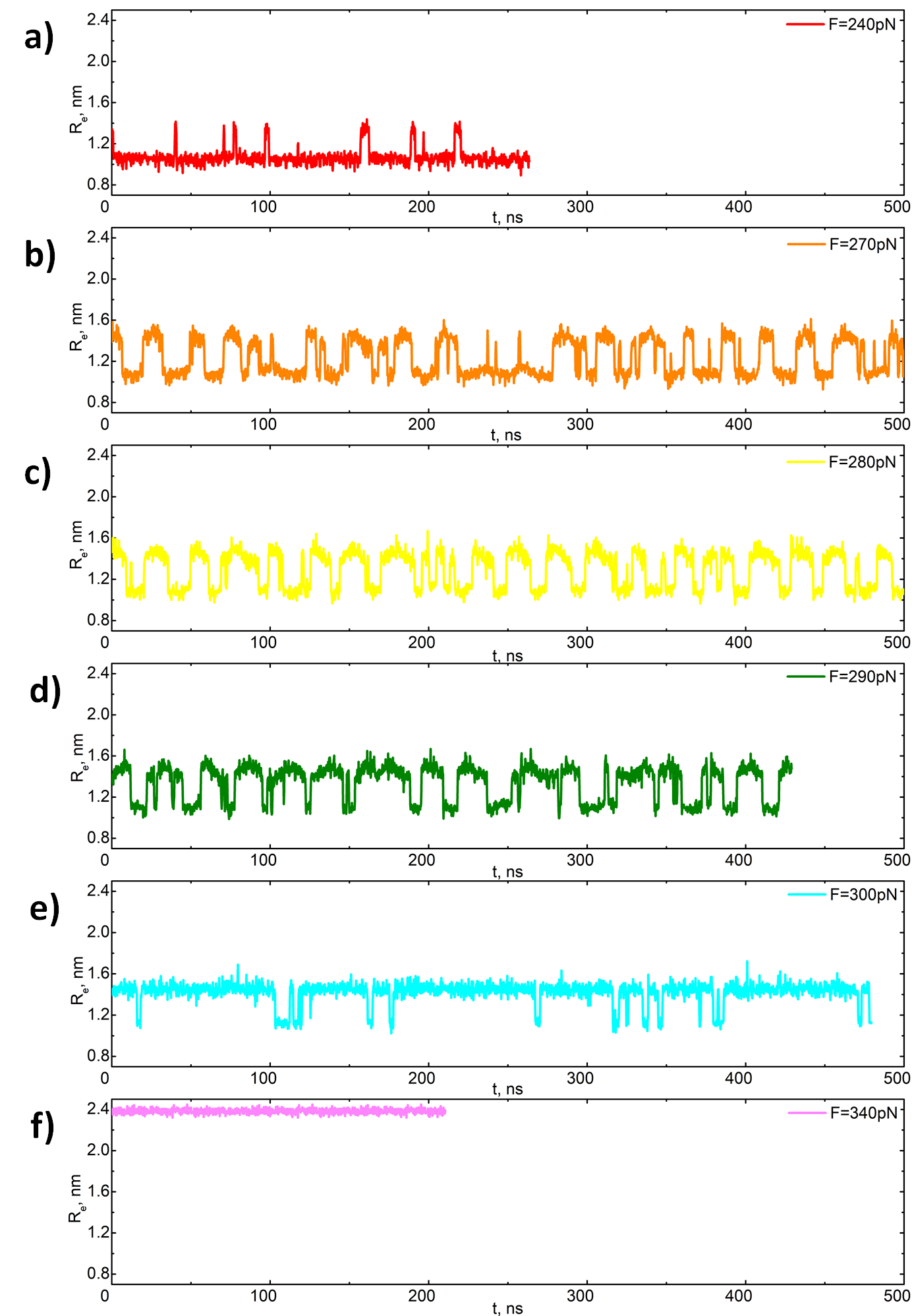}
    \centering
\caption{a-f)Trajectories of spontaneous vibration with different stretching forces. One can see that the oligo-PF5 moves from the close state to the open state with the increasing of the force $F$.}
\label{fig:S3}
\end{figure}
The trajectories of the spontaneous vibrations are shown in Figure \ref{fig:S3}. These trajectories are used to calculate the state distribution, which is shown in Figure \ref{fig:S4}. Optimal conditions for stochastic resonance are obtained from these trajectories. The main condition for clear resonance is a symmetrical distribution of the oligo-PF$5$ states. Under the super-critical load (higher than \SI{320}{\pico\newton} (Figure \ref{fig:S3}f)), the oligo-PF$5$ appears to be in a fully stretched state and is not able to return to the stress-strain state. In the fully stretched state slow transitions from cis-state to trans-state occur and block the sharp transition back.

\subsection{Mean lifetime estimation}
\begin{figure}[H]
    \includegraphics[width=\linewidth]{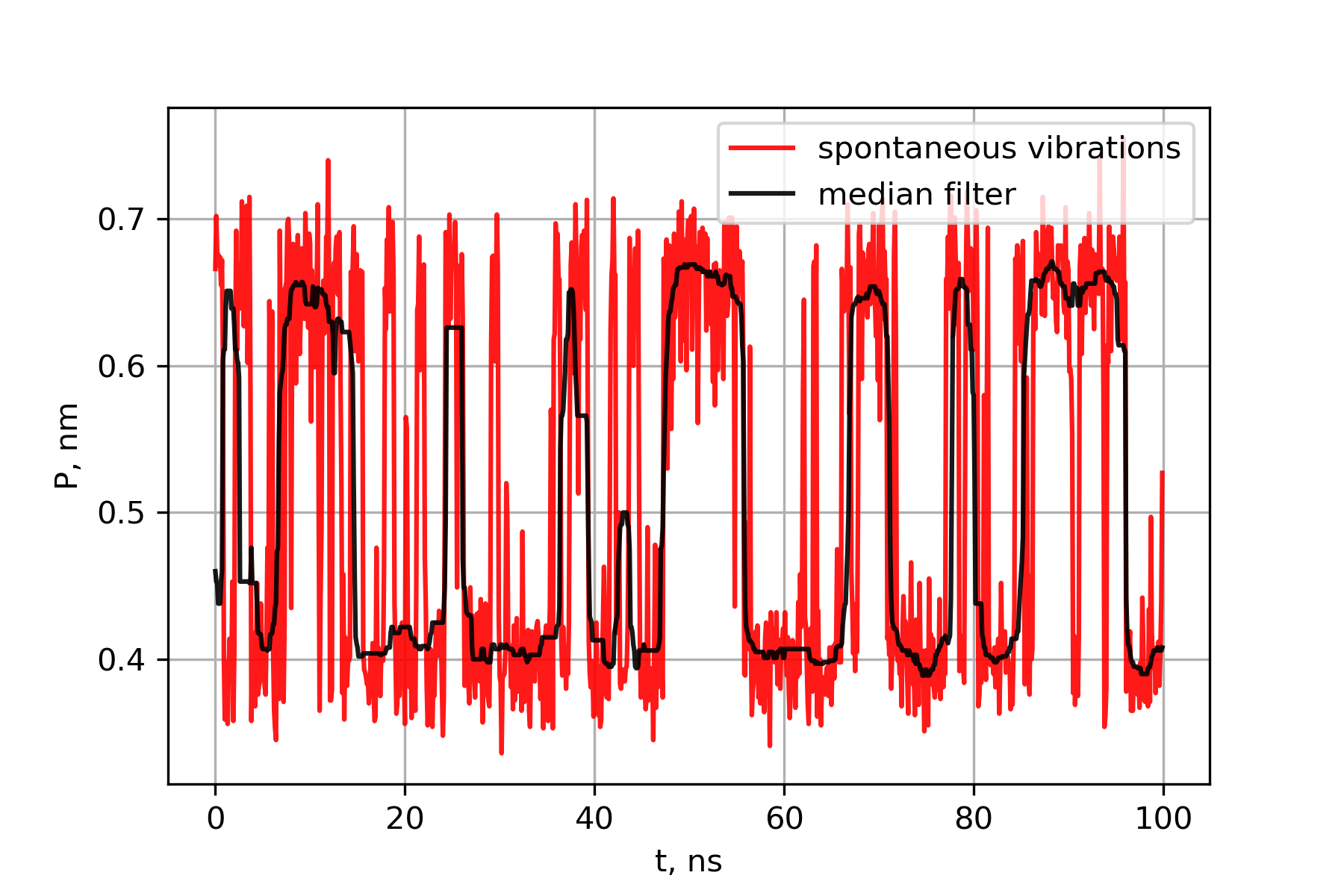}
    \centering
\caption{ The spontaneous vibrations trajectory with and without the median filter usage.}
\label{fig:S5}
\end{figure}
In the oligo-PF systems, one can observe the dynamics at two temporal scales: local and global. The first one is characterized by fast fluctuations near the left-end/right-end states in case of oligo-PF$7$ or squeezed/stress-strain states in case of oligo-PF$5$. However, the global dynamics (the switching between the ends) is more important for stochastic resonance. To calculate the mean lifetime, the local dynamics were excluded using the median filter $ndimage.median\_filter$ from an open-source software SciPy for Python 3 with a median filter window size of \SI{3.5}{\nano\second}, which corresponds to local vibrations. The resulting trajectory is shown in \ref{fig:S5}. According to the estimation, the mean lifetime of oligo-PF$7$ spring was $\tau=\SI{6.5}{\nano\second}$ for the most symmetric distribution at the distance $D=\SI{1.03}{\nano\metre}$ (see Figure 4d). The similar mean lifetime for oligo-PF$5$ was $\tau=\SI{6.14}{\nano\second}$ and corresponded to the external force $F=\SI{279}{\pico\newton}$.

\begin{figure}[H]
    \includegraphics[width=\linewidth]{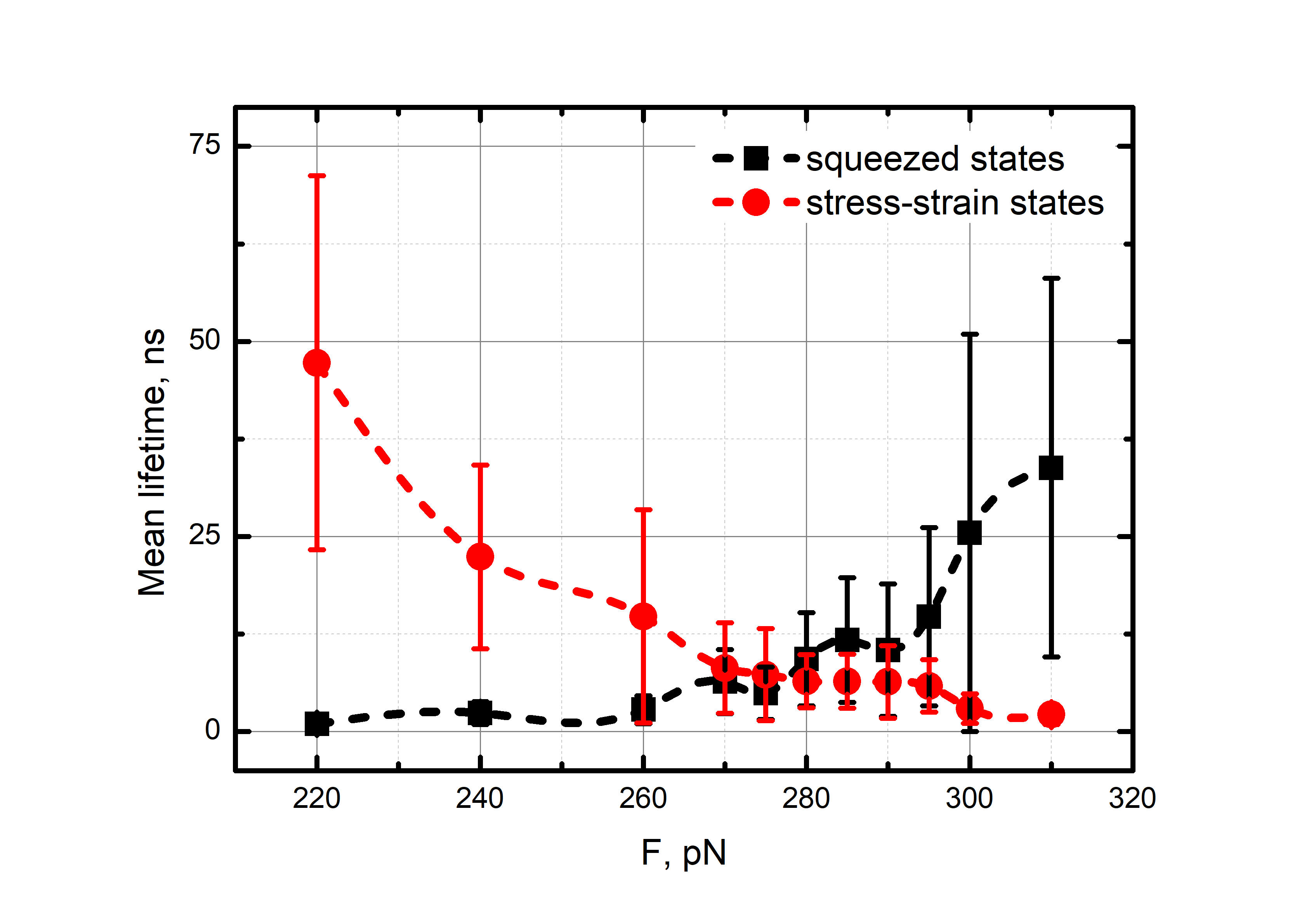}
    \centering
\caption{Mean lifetimes for squeezed (black curve) and stress-strain (red curve) states for the oligo-PF5 are shown.}
\label{fig:S4}
\end{figure}
\subsection{State diagram of oligo-PPF7}
\begin{figure}[H]
    \includegraphics[width=\linewidth]{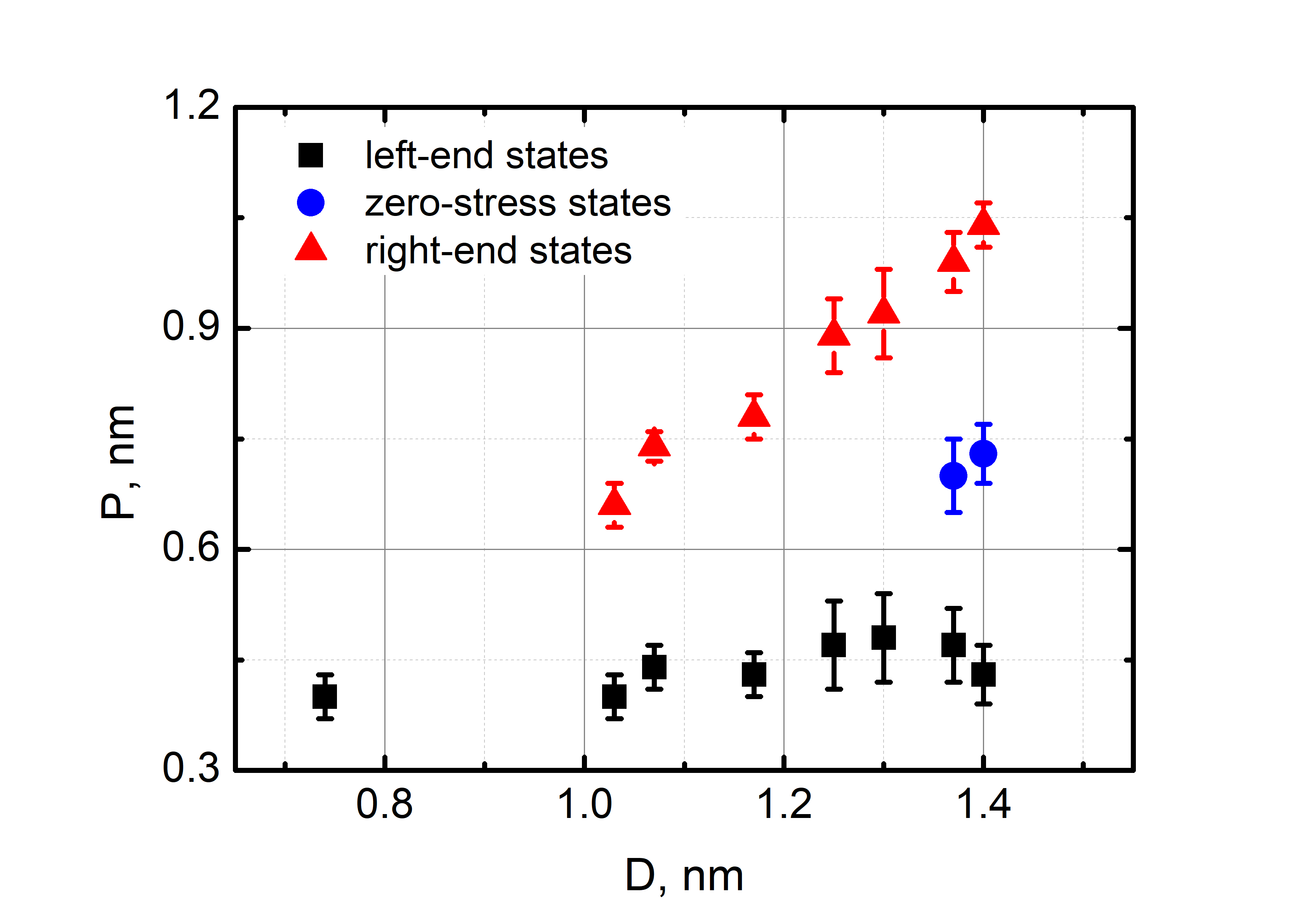}
    \centering
    \caption{The state diagram of oligo-PF7 states.}
    \label{fig:S6}
\end{figure}
The state diagram of the oligo-PF$7$ states with $D>1.30$ nm is shown in \ref{fig:S6}. The wide repulsive zone of the zero-stress states becomes the third attractive state after the threshold of $\sim 4$ stacking lengths. 

\subsection{Signal-to-noise ratio in resonance}
\begin{figure}
\begin{subfigure}{0.49\textwidth}
    \centering
    \includegraphics[width=\linewidth]{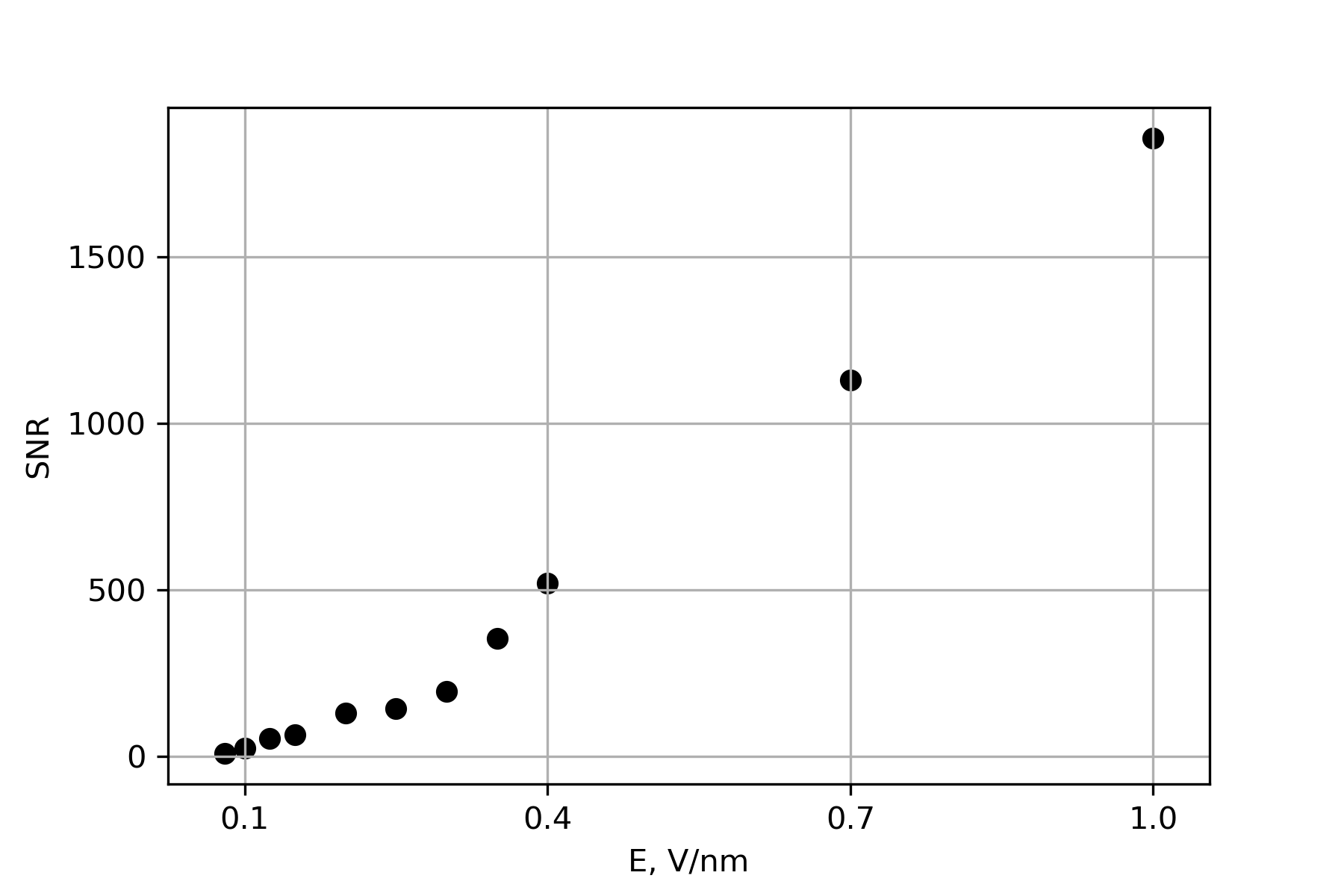}
    \caption{}
\end{subfigure}
\begin{subfigure}{0.49\textwidth}
    \centering
    \includegraphics[width=\linewidth]{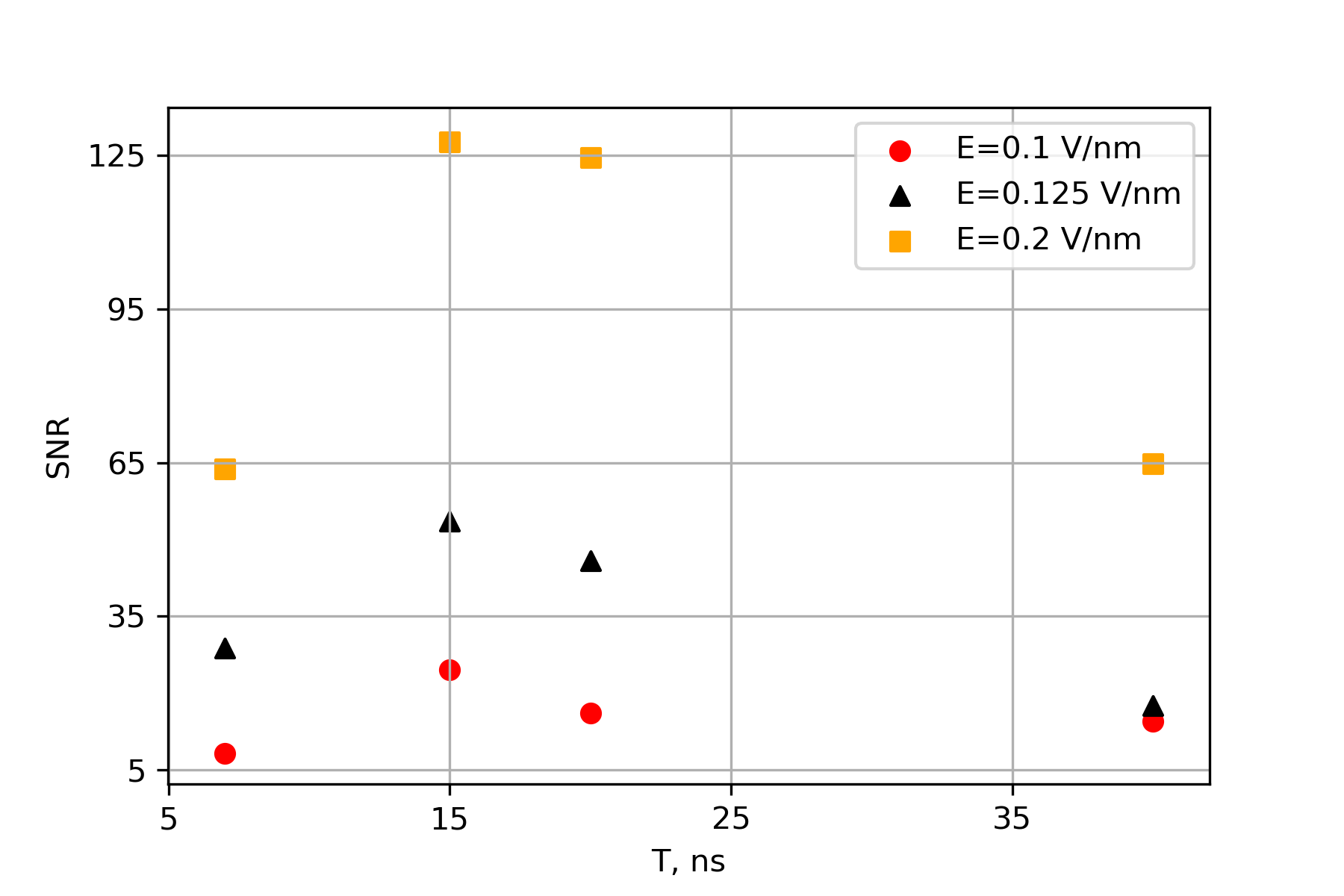}
    \caption{}
\end{subfigure}
    \caption{The signal-to-noise ratio. (a) The dependence of the SNR  on the amplitude E; (b) The dependencies of the SNR on the external oscillation field period with amplitude $E=$0.1, 0.125 and 0.2 V/nm.}
    \label{fig:S7}
\end{figure}
To analyse the stochastic resonance data, power spectra $S(\nu)$ for each trajectory of oligo-PF springs were calculated, which are shown in the Figure 
3b and 5b. The power spectrum is defined as a Fourier transform of an auto-correlation function,
\begin{equation}
\label{eq:S1}
S(\nu)= \int_{- \infty}^{ \infty} \langle  {Z}(t) {Z}(t- \tau) \rangle e^{-2 \pi i \nu \tau} d \tau,
\end{equation}
where $\langle  {Z}(t) {Z}(t- \tau) \rangle$ is the normalized auto-correlation function, since $Z(t)=z(t)/\|z(t)\|$ is the normalised signal of the standard score $z(t)$ for  the oligo-PF$5$ resonance  trajectory $R_e$ (or $P$ in the oligo-PF$7$ case).

The signal-to-noise ratio (SNR) is a standard characteristic of the stochastic resonance. We define SNR as a ratio of the main resonance peak amplitude from the power spectrum of stochastic resonance to the mean noise value near the peak. The dependence of the SNR on the amplitude E is shown on \ref{fig:S7}a. The values could be approximated by two linear series that that intersect at $E=$ 0.3 V/nm, which could mean a transition from stochastic resonance to forced oscillations.

The dependencies of the SNR on the period of oscillating field for different amplitudes $E$ are shown on \ref{fig:S7}b. The main SNR peak is observed close to the doubled mean lifetime of the state in the spontaneous vibration mode. 


\bibliography{references}